\documentclass{pasa}%

\usepackage{graphicx}
\usepackage{subcaption}
\usepackage{placeins}

\usepackage{mathtools} 
\usepackage{bm} 

\usepackage{newtxtext,newtxmath}

\usepackage[Symbol]{upgreek}

\graphicspath{{./figs/}}
\DeclareGraphicsExtensions{.pdf,.png}

\newcommand{\degr}{\hbox{$^\circ$}}
\newcommand{\kpc}{\,\textrm{kpc}} 
\newcommand{\myr}{\,\textrm{Myr}} 
\newcommand{\watts}{\,\textrm{W}} 

\renewcommand{\vec}[1]{\bm{#1}}

\title[CosmoDRAGoN I]{CosmoDRAGoN simulations -- I. Dynamics and observable signatures of radio jets in cosmological environments}

\author[Patrick M. Yates-Jones et al.]{
  Patrick M. Yates-Jones$^{1,2,}$\thanks{E-mail: patrick.yates@utas.edu.au},
  Stanislav S. Shabala$^{1,2}$,
  Chris Power$^{2,3}$,
  Martin G. H. Krause$^{4}$,
  Martin J. Hardcastle$^{4}$,
  Elena A. N. Mohd Noh Velast{\'i}n$^{1}$, and
  Georgia S. C. Stewart$^{1}$\\

  \affil{$^1$School of Natural Sciences, Private Bag 37, University of Tasmania, Hobart, TAS 7001, Australia}%
  \affil{$^2$ARC Centre of Excellence for All Sky Astrophysics in 3 Dimensions (ASTRO 3D)}%
  \affil{$^3$International Centre for Radio Astronomy Research, University of Western Australia, 35 Stirling Highway, Crawley, Western Australia 6009, Australia}%
  \affil{$^4$Centre for Astrophysics Research, University of Hertfordshire, College Lane, Hatfield, Herts AL10 9AB, UK}
}%

\jid{PASA}
\doi{10.1017/pas.\the\year.xxx}
\jyear{\the\year}

\usepackage{aas_macros}
\usepackage[unicode]{hyperref} 
\hypersetup{colorlinks,citecolor=blue,linkcolor=blue,urlcolor=blue,unicode}

\usepackage[noabbrev,capitalise]{cleveref} 

\begin{document}

\begin{frontmatter}
\maketitle

\begin{abstract}
{
  We present the {\bf C}osmological {\bf D}ouble {\bf R}adio {\bf A}ctive {\bf G}alactic {\bf N}uclei (CosmoDRAGoN) project: a large suite of simulated AGN jets in cosmological environments.
  These environments sample the intra-cluster media of galaxy clusters that form in cosmological smooth particle hydrodynamics (SPH) simulations, which we then use as inputs for grid-based hydrodynamic simulations of radio jets.  Initially conical jets are injected with a range of jet powers, speeds (both relativistic and non-relativistic), and opening angles; we follow their collimation and propagation on scales of tens to hundreds of kiloparsecs, and calculate spatially-resolved synthetic radio spectra in post-processing. In this paper, we present a technical overview of the project, and key early science results from six representative simulations which produce radio sources with both core- (Fanaroff-Riley Type I) and edge-brightened (Fanaroff-Riley Type II) radio morphologies. Our simulations highlight the importance of accurate representation of both jets and environments for radio morphology, radio spectra, and feedback the jets provide to their surroundings.
}
\end{abstract}

\end{frontmatter}

\section{Introduction}
\label{sec:cd_intro}
  
  Feedback processes are key to regulating galaxy formation and evolution \citep{Vogelsberger2013,Somerville2015}.
  Typically, both stellar and Active Galactic Nucleus (AGN) feedback are invoked to regulate star formation in both semi-analytic and numerical \citep{CrotonEA06,ShabalaAlexander09,Vogelsberger2014,Schaye2015,LagosEA18,WeinbergerEA18,RaoufEA19,DuboisEA21} galaxy formation models.
  However, only AGN can plausibly provide the energy required to offset runaway cooling in massive ellipticals and galaxy clusters, and subsequent star formation at late cosmological epochs \citep{silk.rees.1998,silk.2005,mcnamara.nulsen.2007}.
  Individual system \citep{BohringerEA93,FabianEA03,FormanEA05} and population studies \citep{SadlerEA89,Burns90,Rafferty2006,MittalEA09} indicate that radio jets (i.e. collimated beams of ionized plasma expelled from near the nuclear black hole and visible at radio wavelengths) are overwhelmingly present in rapidly cooling, massive systems -- precisely where they are needed.
  Moreover, the energy budget \citep{BestEA06,BestEA07,TurnerShabala15,HardcastleEA19} and duty cycle \citep{BestEA05,PopeEA12,SabaterEA19} of jet activity strongly suggest that the majority of AGN jets operate as cosmic thermostats \citep{Kaiser2003}, with rates of energy input likely balanced on average over long timescales by the cooling of hot gas atmospheres \citep{Vernaleo2007,Yang2016,Martizzi2019}.
  Because of this, so-called ``maintenance mode'' feedback is a key feature of all galaxy formation models.
  
  Exactly how and where jets impact their host galaxy environments through feedback has been the subject of many numerical jet simulations \citep[e.g.][]{Zanni2005,Hardcastle2014,Mukherjee2018a,Bourne2021}.
  Yet except for a small number of studies \citep{Heinz2006,Morsony2010,Mendygral2012,Bourne2021}, the description of the host galaxy environment has been relatively simple in comparison to the kinds of dynamic environment found in cosmological simulations.
  Some theoretical studies incorporate both jet-inflated lobes and complex environments \citep[e.g.][]{Ehlert2021,Vazza2021}, but do not have sufficient resolution to model the sub-kpc jet physics responsible for the production of the large-scale radio lobes in the first place.
  Cosmological simulations \citep[see, e.g.][]{Dubois2014,Vogelsberger2014,Schaye2015,Cui2018,Lee2021} capture complicated galaxy group and/or cluster dynamics, but for computational reasons are limited to comparatively simple models of jets; these are commonly incorporated as heavy, slow outflows.
  On the other hand, simulations have shown the importance that light relativistic \citep{Saxton2002,Zanni2003,Krause2003,Krause05,English2016,Perucho2019b} and initially conical \citep{Krause2012} jets have for large scale morphology.
  There is therefore a clear need for more sophisticated simulations of jets in dynamic environments.

  Observational and theoretical evidence suggests that observed radio source properties are strongly influenced by their host environments \citep{Hardcastle2014,Rodman2019,Lan2021,YatesJonesEA22}.
  Large source samples exhibiting complex jet dynamics are increasingly being observed in new radio surveys such as LOFAR LoTSS \citep{Shimwell2017,ShimwellEA22}, ASKAP EMU \citep{NorrisEA21}, and the VLA Sky Survey \citep{LacyEA20}, thanks to their increased sensitivity to low-power jets  -- precisely the structures that are more susceptible to environmental effects because of buoyancy \citep{Saxton2001,Krause2012}.
  Numerical simulations of jet dynamics, together with an appropriate framework for calculating the synthetic radio emission, are required to interpret these observational data.
  This demands accurate treatment of particle acceleration and loss mechanisms coupled to the jet dynamics. Such an approach is essential for connecting the observable properties of radio jet populations – namely synchrotron radio emission – to the location and magnitude of the feedback they provide.\\
  
  In this paper, we introduce the {\bf C}osmological {\bf D}ouble {\bf R}adio {\bf A}ctive {\bf G}alactic {\bf N}uclei (CosmoDRAGoN) simulation suite, which aims to tackle the above questions by embedding sophisticated dynamical simulations of 
  jets into realistic environments derived from cosmological hydrodynamical simulations, and exploring a broad range of jet and environmental parameters.
  We use environments from cosmological hydrodynamical galaxy formation simulations of  individual galaxy clusters in \textsc{the three hundred} project \citep{Cui2018}.
  Our simulated jets are conical and relativistic\citep{YatesJonesEA21}, and we adopt a detailed treatment of electron acceleration and loss processes to calculate synthetic radio emission \citep{YatesJonesEA22}.
  Several authors \citep[e.g.][]{jones.1999,tregillis.2001} have calculated the details of shock acceleration and ageing numerically within magneto-hydrodynamic simulations.
  Our method, detailed in \citet{YatesJonesEA22} employs a more flexible semi-analytic approach: we record the simulated dynamics for tracer particles representing packets of electrons to quantify the sites of particle acceleration at shocks and losses due to source expansion. With this saved information we calculate synchrotron and Inverse Compton losses in post-processing.
  In this way, we are able to efficiently cover a broad range of parameter space within a single simulation, including redshifts and (not well constrained) lobe magnetic field strengths.
  
  The rest of the paper proceeds as follows.
  In \cref{sec:cd_simulations} we present the simulation method, initial condition generation, the post-processing procedure and data outputs, and the parameter space covered.
  In \cref{sec:cd_results} we present early science results using a subset of simulations from the full CosmoDRAGoN suite. Our jet simulations are capable of producing both edge-brightened, Fanaroff-Riley Type II \citep[FR II]{FanaroffRiley74} and core-brightened, Fanaroff-Riley Type I \citep[FR I]{FanaroffRiley74} radio source morphologies; we discuss these in \cref{sec:cd_frII} and \cref{sec:cd_frI}, respectively.
  We discuss our results in \cref{sec:cd_discussion} and then conclude with a summary of the CosmoDRAGoN simulation suite and future outlook in \cref{sec:cd_summary}.

\section{Simulations}
\label{sec:cd_simulations}

  \begin{figure*}
    \centering
    \includegraphics{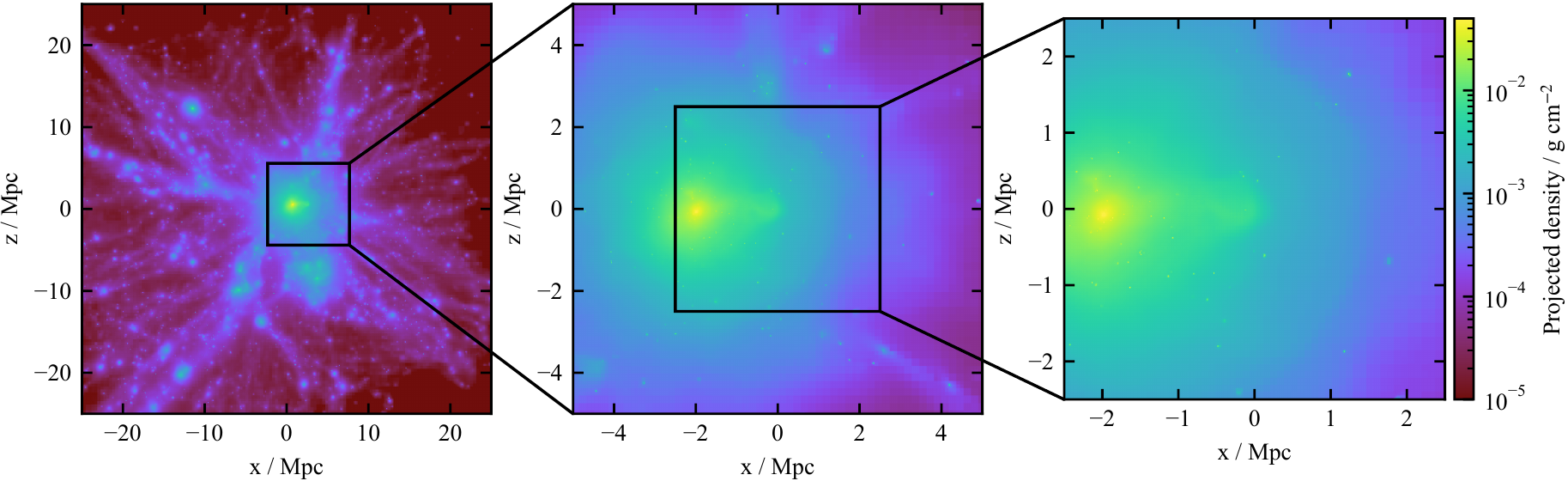}
    \caption{
      Projected density maps of CosmoDRAGoN environment \textit{002-0003}.
      \textit{Left panel:} The full cluster from Three Hundred project (cluster 002).
      \textit{Middle and right panels:} Zoom-ins centred on subhalo 0003.
    }
    \label{fig:cd_example_cluster}
  \end{figure*}

  \begin{figure*}
    \centering
    \includegraphics{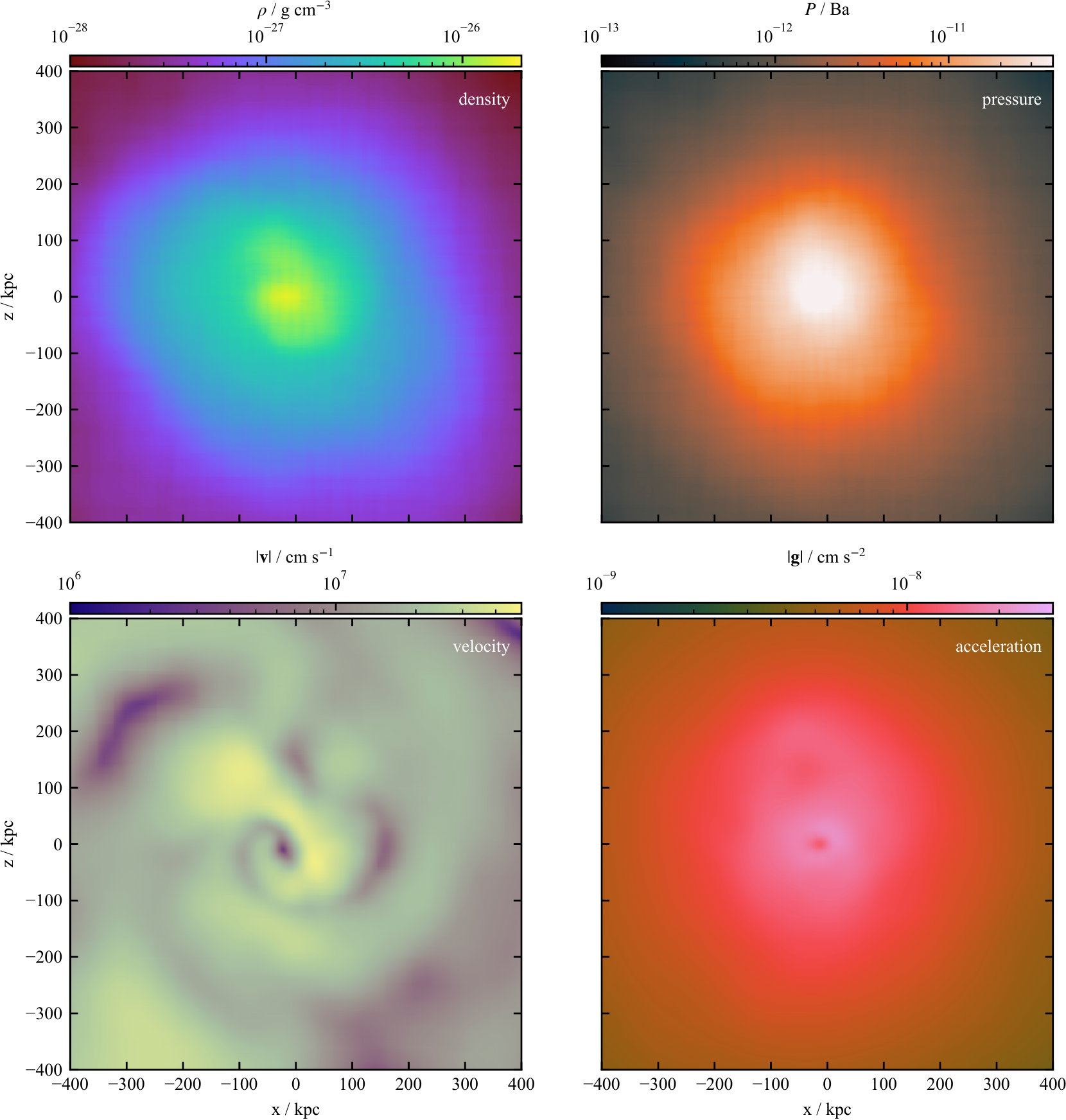}
    \caption{
      Environment \textit{002-0003} quantities after interpolation onto a regular three-dimensional Cartesian grid.
      Midplane slices at $y=0$ of (left to right, top to bottom): density, pressure, velocity magnitude, gravitational acceleration magnitude.
    }
    \label{fig:cd_example_cluster_interpolated}
  \end{figure*}

  \begin{figure}
    \includegraphics{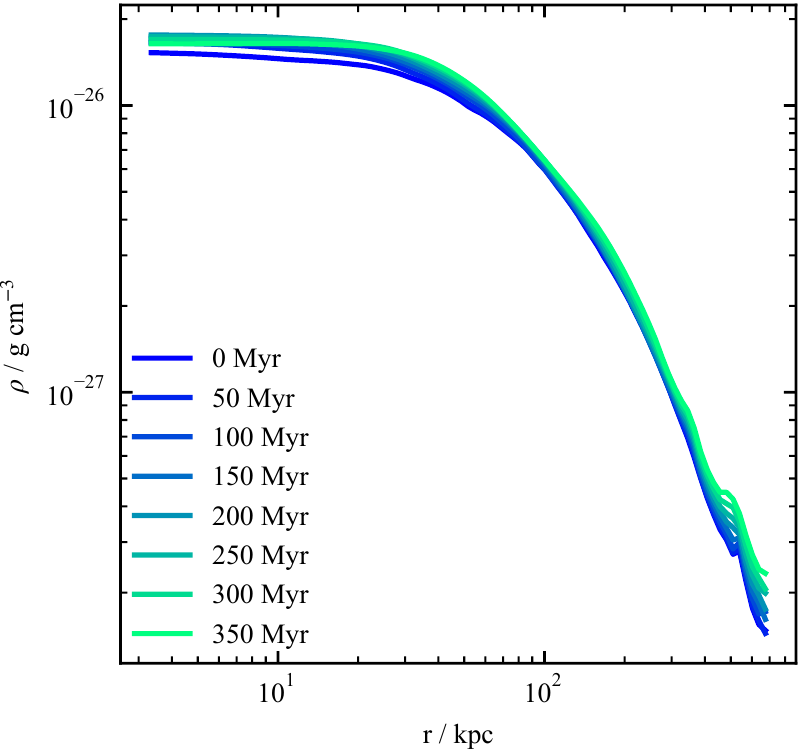}
    \caption{
      Time evolution of the average radial density profiles for environment \textit{002-0003}.
      The environment is evolved without a jet for $350\myr$.
    }
    \label{fig:cd_example_cluster_stability}
  \end{figure}

  \begin{table}
    \caption{
      Parameters of the six representative simulations.
      $Q$ is the total one-sided jet power of the radio source.
      $v_\textrm{j}$ is the initial jet velocity,
      and $\theta_\textrm{j}$ is the half-opening angle.
    }
    \centering
    \label{tbl:cd_jet_population_parameters}
    \begin{tabular}{ l r r r r }
      \hline
      Code & $Q$ & $v_\textrm{j}$ & $\theta_\textrm{j}$ & Morphology \\
           & (W) & ($c$) & ($\degr$) & \\
           \hline \\ [-0.75em]
      Q38-v98-$\uptheta$7.5 & $10^{38}$ & $0.98$ & $7.5\degr$ & FR II \\
      Q38-v98-$\uptheta$25 & $10^{38}$ & $0.98$ & $25\degr$ & FR II \\
      Q38-v30-$\uptheta$25 & $10^{38}$ & $0.3$ & $25\degr$ & FR II \\
      Q36-v30-$\uptheta$25 & $10^{36}$ & $0.3$ & $25\degr$ & FR II \\
      Q36-v01-$\uptheta$25 & $10^{36}$ & $0.01$ & $25\degr$ & FR I \\
      Q36-v01-$\uptheta$30 & $10^{36}$ & $0.01$ & $30\degr$ & FR I \\
      \hline
    \end{tabular}
  \end{table}
  
  The simulations presented here combine grid-based hydrodynamical jet models with galaxy cluster environments taken from \textsc{the three hundred} project cosmological hydrodynamical simulations \citep{Cui2018}.
  In this section, we explain how these are combined in the CosmoDRAGoN setup.
  First, we describe the numerical techniques (\cref{sec:cd_numerical_setup}) and jet injection model (\cref{sec:cd_jet_injection}) used.
  Next, we describe the cosmological environments and their conversion to initial conditions (\cref{sec:cd_initial_conditions}), then detail how we validate their stability (\cref{sec:cd_environment_stability}).
  We conclude this section by discussing the simulation suite parameters (\cref{sec:cd_simulation_suite}) and data products (\cref{sec:cd_output_data}).

  \subsection{Numerical setup}
  \label{sec:cd_numerical_setup}

    The simulations in the CosmoDRAGoN project are carried out using a modified version of \textsc{pluto}\footnote{\url{http://plutocode.ph.unito.it/}} 4.3, a freely available grid-based simulation code developed for high Mach number astrophysical fluid flows \citep{Mignone2007}.
    \textsc{pluto} supports several different physics modules; in this work, we use the relativistic hydrodynamic physics module.
    The fluid is evolved on a static three-dimensional Cartesian grid by solving the conservation laws using the HLLC Riemann solver with linear reconstruction.
    2$^\textrm{nd}$ order Runge-Kutta time-stepping is used to advance the simulation in time, with a Courant-Friedrichs-Lewy (CFL) number of 0.3.
    To increase the simulation robustness, we make use of the shock flattening feature in \textsc{pluto} to switch to the HLL solver and the \texttt{MINMOD} limiter in the presence of strong shocks.
    The gas pressure is recovered using entropy by default (for which a separate conservation equation is solved), however total energy is used in the presence of strong shocks.
    The Taub-Mathews equation of state \citep{Taub1948,Mathews1971,Mignone2007a} is used to model the thermodynamic evolution of the simulation.
    This equation of state models fluids that are either non-relativistic, ultra-relativistic, or somewhere in between; both the environment and jet thermodynamics are well modelled by this.
    Cooling will be included in a subset of the final simulation suite using a tabulated method with non-equilibrium cooling rates determined by the \textsc{mappings~v} code \citep{Sutherland2018}; however, the simulations presented in detail here do not simulate cooling of the hot gas nor radiative losses of the relativistic plasma.

    The standard $\Lambda$CDM cosmology is used throughout this project for relating physical and observable quantities, with parameters obtained from the \textit{Planck} mission \citep{Planck2016}: $\Omega_M=0.307$, $\Omega_B=0.048$, $\Omega_\Lambda=0.693$, $h=0.678$.
    These values are consistent with the cosmological simulation catalogue \citep{Cui2018} from which our initial conditions are derived (discussed further in \cref{sec:cd_initial_conditions}).

    CosmoDRAGoN simulations are carried out on a static three-dimensional Cartesian grid.
    This grid is defined as per-coordinate patches of varying resolution (i.e. the grid
    size varies independently in each coordinate), to maximise computational efficiency.
    The jet injection patch, which stretches from $-2.5$ to $2.5\kpc$, is uniformly covered by 100 cells along each dimension for a resolution of $0.05\kpc / \textrm{cell}$.
    This high-resolution patch ensures that jet injection and collimation are resolved with several cells across the jet beam, which is sufficient to correctly capture the collimation dynamics\footnote{Because the jet beam radius at collimation depends on a range of factors, including environment, this resolution has been deduced by running
    several simulations and examining how the jet collimates.}.

    The rest of the simulation grid is covered by geometrically stretched patches out to the grid boundaries, such that the edges of the simulation grid have the coarsest resolution. The grid is stretched in each coordinate from $2.5~(-2.5)~\kpc$ to 
    $10~(-10)~\kpc$ over 100 grid cells, and then from $10~(-10)~\kpc$  to $200~(-200)~\kpc$ over 330 grid cells. The actual spacing is determined internally
    by PLUTO according to
    \begin{equation}
        r\frac{1-r^N}{1-r}=\frac{x_R-x_L}{\Delta\,x},
    \end{equation}
    where $r$ is the stretching ratio, $\Delta\,x$ is taken from the closest uniform grid, $N$ is the number of points in the stretched grid, and $x_L, x_R$ are the left- and
    rightmost points of the patch. The simulation grid has a typical resolution of $0.10\kpc / \textrm{cell}$ at $10\kpc$ and $0.85\kpc / \textrm{cell}$ at $100\kpc$.

    Outflow boundary conditions are enforced at the grid boundaries, setting the gradient of simulated quantities to $0$ across the boundary.
    These boundary conditions favour the cosmological environments used here by dampening any residual environment oscillations, rather than amplifying them as in the case of reflective boundaries (cf. \cref{sec:cd_three_hundred}).

    The CosmoDRAGoN simulations are run on the \textit{Gadi} facility provided by the National Computational Infrastructure, Australia.
    Each simulation runs on up to 4608 Intel Xeon 8274 processors, with parallelisation using the MPI specification.
    While the bulk of the analysis is also run on \textit{Gadi}, some supplementary simulations and analysis make use of the \textit{kunanyi} facility, provided by the Tasmanian Partnership for Advanced Computing.

  \subsection{Jet injection}
  \label{sec:cd_jet_injection}

    The injected jets are modelled as conical outflows with a half-opening angle $\theta_\textrm{j}$ from a spherical injection zone; this is the same injection model used in our previous work \citep{YatesJonesEA21}.
    The injection zone is defined as a sphere with radius $r_0 = 0.75\kpc$, centred at the origin; this is appropriate for conical jets without a dynamically important magnetic field,
    as is the case in our work.
    Within this injection zone, the fluid quantities are continuously updated with the jet injection values.
    For a given desired jet density, $\rho_\textrm{j}$, and jet pressure, $P_\textrm{j}$ at $r_0$, the injection zone values are calculated throughout the injection sphere as
    \begin{align}
      \rho_\textrm{i}(r) &= 2 \rho_\textrm{j} (1 + (r / r_0)^2)^{-1}\\
      P_\textrm{i}(r) &= 2^\Gamma P_j \left(\frac{\rho(r)}{\rho(r_0)}\right)^\Gamma
      \,,
    \end{align}
    where $\Gamma$ is the adiabatic index and $r$ is the spherical radius with respect to the origin.
    The velocity is defined radially outwards from the origin as $v_r = v_\textrm{j}$ if $\theta \le \theta_\textrm{j}$, and $\vec{v} = 0$ elsewhere.
    Additionally, we inject a jet tracer fluid with an initial value of $1.0$ if $\theta \le \theta_\textrm{j}$, and $0.0$ elsewhere.

    The one-sided relativistic jet power\footnote{The one-sided jet power is the rate of energy injection from one of a pair of anti-parallel jets; for each of the jets,
    the one-sided power is the sum of the kinetic and thermal energy components.}is given as
    \begin{equation}
      Q = \left[ \gamma (\gamma - 1) c^2 \rho_\textrm{j} + \gamma^2 \frac{\Gamma}{\Gamma - 1} P_\textrm{j} \right] v_\textrm{j} A_\textrm{j}\,,
      \label{eqn:cd_rel_jet_power}
    \end{equation}
    with the speed of light in a vacuum $c$, bulk flow Lorentz factor $\gamma = 1 / \sqrt{1 - v_\textrm{j}^2/c^2}$, and cross-sectional area of the jet inlet $A_\textrm{j}$.
    For a given jet velocity $v_\textrm{j}$ and area $A_\textrm{j}$, the temperature parameter of the jet plasma $\Theta = P_\textrm{j} / (\rho_\textrm{j} c^2)$ \citep{Mignone2007a} uniquely defines the jet density and pressure.
    We restrict our focus to the injection of cold jets, $\Theta \ll 1$, and so use the ideal equation of state with $\Gamma = 5/3$ to calculate the initial jet properties.
    At our injection radius, adiabatic expansion would have dissipated any significant pressure a jet might have had close to its initial formation site.
    Any re-heating via interaction with the environment is taken into account as far as it is explicitly modelled in our hydrodynamic simulations.
    For non-relativistic jets, the one-sided jet power reduces to simply the flux of kinetic energy density along the jet,
    \begin{equation}
      Q = \frac{1}{2} \rho_\textrm{j} v_\textrm{j}^3 A_\textrm{j} \,,
      \label{eqn:cd_nonrel_jet_power}
    \end{equation}

    The Lagrangian particle module in \textsc{pluto} is used to inject tracer particles\footnote{These particles are used to track jet backflow and shocks with significantly higher temporal resolution than can be achieved with solely grid outputs, due to the much smaller file size.} into the 
    jet injection zone. Synthetic synchrotron emission is calculated per particle; each particle is taken to represent an ensemble of electrons.
    To do this, we use the PRAiSE framework presented in \citet{YatesJonesEA22} to evolve the electron energy distribution in time, including both radiative and adiabatic losses.
    All emissivities are calculated in post-processing using a Voronoi tesselation to assign appropriate volumes to each tracer particle, thus allowing us to probe observable properties across a range of parameters (e.g. frequency, redshift) without the need to rerun a simulation, avoiding significant computational expense.
  
  \subsection{Initial conditions}
  \label{sec:cd_initial_conditions}

    \subsubsection{The Three Hundred Project}
    \label{sec:cd_three_hundred}

      Environments taken from cosmological simulations are a defining feature of the CosmoDRAGoN simulations.
      We draw our environments from \textsc{the three hundred} project \citep{Cui2018}, a suite of 324 cosmological zoom simulations of galaxy clusters run with full galaxy formation physics.
      These were identified as the most massive galaxy clusters in the dark matter only MultiDark simulation \citep[MDPL2]{Klypin2016} and resimulated with a range of astrophysics codes.
      The simulated clusters used in this work were run with \textsc{gadget-x} \citep{Beck2016}, a variant of the \textsc{Gadget2} code of \citep{springel2005} that incorporates an improved implementation of smoothed particle hydrodynamics (SPH).
      In addition, \textsc{gadget-x} includes a range of physical prescriptions to model radiative cooling, star formation, black hole growth, and stellar and AGN feedback.
      Further details can be found in \citet{Cui2018}.
      We note that, while the implementation of AGN feedback in \textsc{the three hundred} clusters is less realistic than in CosmoDRAGoN jet simulations -- a necessity due to the large dynamic range of scales samples by the cosmological simulations -- when averaged over timescales of hundreds of Myr representative of the typical time between \textsc{the three hundred} snapshots, this implementation provides the right level of feedback to produce the realistic environments required for CosmoDRAGoN simulations.
      
      Particle data is stored in 128 snapshots, equally spaced in the natural logarithm of the expansion factor between redshifts $z=17$ to $z=0$, and halo catalogues are generated for each snapshot using \textsc{ahf}\footnote{\url{http://popia.ft.uam.es/AHF}} halo finder \citep{Knollmann2009}.
      In this paper, we use outputs at $z=0$, however, the full CosmoDRAGoN suite will feature environments at a range of redshifts. We follow \citet{Cui2018} in their classification of dynamical state and focus on relaxed clusters (cf. \cref{sec:cd_environment_selection}).

    \subsubsection{Creating a realistic cosmological environment}
    \label{sec:cd_interpolation}
      We wish to model jet propagation in a background, defined as a 3-dimensional mesh, whose properties (e.g. density, pressure, momentum) closely match those of the simulated clusters.
      An important requirement for this work is that the environment is stable.
      Because \textsc{pluto} does not support self-gravity, we calculate the gravitational acceleration using \textsc{gadget-2} and interpolate it onto the mesh along with the other quantities (e.g gas density, pressure, momentum), as described below. 
      We note that \textsc{pluto} cannot follow the evolution of the gravitational potential, and so this limits the maximum time over which an environment remains stable before gas motions present within the simulated cluster lead to a disassociation between the gas and corresponding gravitational potential.
      We investigate this limitation in \cref{sec:cd_environment_stability}.
    
      To smooth the simulated cluster quantities onto the 3-dimensional mesh, we use the standard `scatter' formalism for SPH interpolation, where the smoothed quantity $A_s$ as a function of position $\vec{r}$ is given by a summation over $i$ particles as
      \begin{equation}
        A_s(\vec{r}) \approx \sum_i m_i \frac{A_i}{\rho_i} W(|\vec{r} - \vec{r_i}|, h_i)\,.
        \label{eqn:cd_sph_interp}
      \end{equation}
      Here $m_i$ is particle mass, $\rho_i$ is particle density, $h_i$ is particle smoothing length, and $W(\vec{r}, h)$ is the smoothing function.
      This interpolation approach conserves total mass in a smoothed field, $\int \rho_s(\vec{r}) d\vec{r} = \sum_i m_i$.
      We use a modified version of \textsc{sphtool}\footnote{\url{https://bitbucket.org/at_juhasz/sphtool}} to perform the actual interpolation onto a Cartesian grid with $1\kpc/\textrm{cell}$ resolution.
      We adopt the cubic spline \citep[or $M_4$ kernel, ][]{Monaghan1985} for the interpolation process, giving the smoothing function as $W(r,h)=M_4(r)/h^3$.

      The particle density, pressure, momentum density and force density are interpolated using \cref{eqn:cd_sph_interp}.
      Post-interpolation, the velocity and acceleration fields are recovered from the momentum and force fields respectively.
      The velocity field is corrected for the bulk velocity of the environment before interpolation.
      The resulting interpolated environment quantities are suitable for loading into \textsc{pluto} as initial conditions.
      We note that the interpolation grid does not necessarily match the simulation grid used in \textsc{pluto}; initial conditions are interpolated onto the simulation grid using tri-linear interpolation.
   
    \subsubsection{Environment selection}
    \label{sec:cd_environment_selection}
      We begin by identifying massive, dynamically relaxed, clusters that have had no recent major mergers.
      Using the halo catalogue for a selected cluster, the most massive subhalos are identified and visually inspected to verify that they are not involved in a significant merger event at the epoch of interest ($z=0$ for this work).
      The degree of hydrostatic equilibrium of the cluster is calculated and visually inspected to ensure there are no large unstable areas.
      Next, the fraction of the cluster that is in hydrostatic equilibrium with the underlying gravitational field is calculated, and clusters significantly out of hydrostatic equilibrium are removed from the list of candidates.
      Finally, each candidate subhalo is interpolated onto a three-dimensional Cartesian grid as in \cref{sec:cd_interpolation}, and the stability of this environment within \textsc{pluto} is tested as described in \cref{sec:cd_environment_stability}.

      In \cref{fig:cd_example_cluster} we show one such subhalo and its parent cluster from the Three Hundred project; this subhalo has been identified as suitable for the CosmoDRAGoN simulations: identified as subhalo 0003 in cluster 002 (with halo mass $M_\textrm{halo}=2.02\times10^{14} \textrm{M}_{\odot}$, virial radius $R_\textrm{halo}=1.24\,\textrm{Mpc}$, and central density $\rho=1.53\times10^{-26}\,\textrm{g cm}^{-3}$), it is given the code \textit{002-0003}.
      The corresponding interpolated quantities are shown in \cref{fig:cd_example_cluster_interpolated}.

  \subsection{Environment stability}
  \label{sec:cd_environment_stability}

    The CosmoDRAGoN simulations do not evolve the gravitational potential due to gas self-gravity with time, nor are any dark matter particles included.
    This is sufficient for our focus on jet morphology and evolution: a static gravitational field is appropriate provided the environment remains reasonably stable over a typical jet active and remnant lifetime of up to a few hundred Myrs.

    The relative stability of each suitable subhalo identified using the methods in \cref{sec:cd_environment_selection} is confirmed by evolving the environment in \textsc{pluto} (with no jet) for $500\myr$; significantly greater than the maximum active plus remnant lifetimes in typical CosmoDRAGoN simulations.
    We require that the radially averaged density and pressure should not vary by more than $0.5-1\,\textrm{dex}$ over this time, and the average per-coordinate velocities should also not exceed $\sim 500\,\textrm{km s}^{-1}$.
    In \cref{fig:cd_example_cluster_stability} we show the change in average density as a function of radius for environment \textit{002-0003} over $350\myr$.
    While there is some evolution in the density profile, including a small inwards-propagating perturbation caused by the grid boundaries, it is small over the simulated time-scale, confirming that this subhalo is suitable for the jet lifetimes simulated in CosmoDRAGoN.
    In addition, the disturbance never gets close to the jet on the simulated timescales: it is $>400\kpc$ from the origin at $350\myr$, while the maximum lobe length is $<150\kpc$ (see \cref{fig:cd_jet_population_density}).

  \subsection{Simulation suite}
  \label{sec:cd_simulation_suite}

    The jet and environment parameters in the CosmoDRAGoN simulation suite are chosen to produce a varied population of radio sources.
    We simulate a range of kinetic jet powers, typical of both low-power (FR I), and medium to high power (FR II) radio sources.
    Several velocities are simulated, ranging from mildly supersonic to strongly relativistic; these cover the observed range of jet velocities \citep{Laing2014,Hardcastle1999}.
    Following \citet{Krause2012}, several jet half-opening angles are considered.
    These range from narrow half-opening angles likely to produce FR II morphology ($\theta_\textrm{j}=7.5\degr$) to wide half-opening angles likely to produce FR I morphology ($\theta_\textrm{j} \ge 25\degr$) after a sufficient length of time.

    Several cosmological environments are used as initial conditions, covering both poor groups and clusters; specifically, in this work we consider a galaxy group with
    virial mass and radius of $1.9 \times 10^{13} \textrm{M}_{\odot}$ and $0.565~\textrm{Mpc}$, and a cluster with $2 \times 10^{14} \textrm{M}_{\odot}$ and $1.2~\textrm{Mpc}$.
    The initial environment velocity is zeroed for the simulations presented in this paper, while a greater variety of initial conditions will be explored in the full simulation suite.
    Our simulations use a static injection region, although we note that 
    moving jet injection regions are a possible cause behind both wide- and narrow-angle tailed observed radio source morphology \citep{ONeill2019}.
    Observed radio jets are likely to have complex outburst histories including both active and remnant phases \citep{Shabala2008,Brienza2017,ShabalaEA20,Morganti2021}; to this end, we simulate both phases of jet evolution.

  \subsection{Data products}
  \label{sec:cd_output_data}

    The primary data products from CosmoDRAGoN are \textsc{pluto} grid and particle data files.
    The grid data files are output with a temporal resolution of $2\myr$ or better and contain the values of density, pressure, velocity, and a tracer for each grid cell.
    The electron-packet-tracing particle data files are output with a temporal resolution of $0.1\myr$ or better, and contain for each particle in the simulation its coordinates, velocity, injection time, tracer value, and last shocked time \citep[for three shock thresholds $\epsilon_p=0.05, 0.5, 5.0$, see][]{YatesJonesEA22}, density, and pressure.
    The particles are assigned grid values at each timestep using a triangular-shaped cloud interpolation.
    The grid data files are compressed using \textsc{zfp} compression \citep{Lindstrom2014}, for a compression factor of $\sim4$x.

    A processing pipeline has been developed to produce reduced data outputs.
    This automated pipeline produces slices and projections of the grid quantities, calculates jet dynamic information (length and volume), and calculates particle emissivity for a given set of observing parameters.
    Once these quantities have been calculated, the pipeline produces diagnostic plots of the simulations which are used to verify their accuracy.

\section{Results}
\label{sec:cd_results}

  The CosmoDRAGoN simulation suite produces radio sources with a variety of morphologies and probes a range of feedback effects, reflecting the jet parameters and the environments into which the jets propagate.
  In the following sections, we look at six representative simulations that are likely to produce FR II (\cref{sec:cd_frII}) and FR I (\cref{sec:cd_frI}) morphologies.
  The parameters for these simulations are listed in \cref{tbl:cd_jet_population_parameters}, along with the morphology we would expect given the choice of kinetic jet power, initial jet velocity, and half-opening angle.
  The parameter space of these simulations covers both the high and low one-sided jet powers ($Q=10^{36},10^{38}\watts$), small and large jet half-opening angles ($\theta_\textrm{j} = 7.5\degr, 25\degr$), and three velocities ($v_\textrm{j}/c = 0.01,0.3,0.98$). The simulations propagate into the \textit{002-0003} cluster-like environment.
  The initial environment velocities are zeroed in the simulations presented here.

  Fast, relativistic, high power jets are expected to produce FR II morphologies, while low power, slower jets (on scales of several kpc) with wider opening angles are expected to produce FR I morphologies \citep{Krause2012,Laing2014}.
  In the next two sections we confirm this to be so, both in terms of jet dynamics and associated radio emission. The spectral index $\alpha$ is defined by $S=\nu^{-\alpha}$ for flux density $S$ and frequency $\nu$ for this paper.

  \subsection{High power radio jets in cosmological environments}
  \label{sec:cd_frII}

    \begin{figure*}
      \centering
      \includegraphics{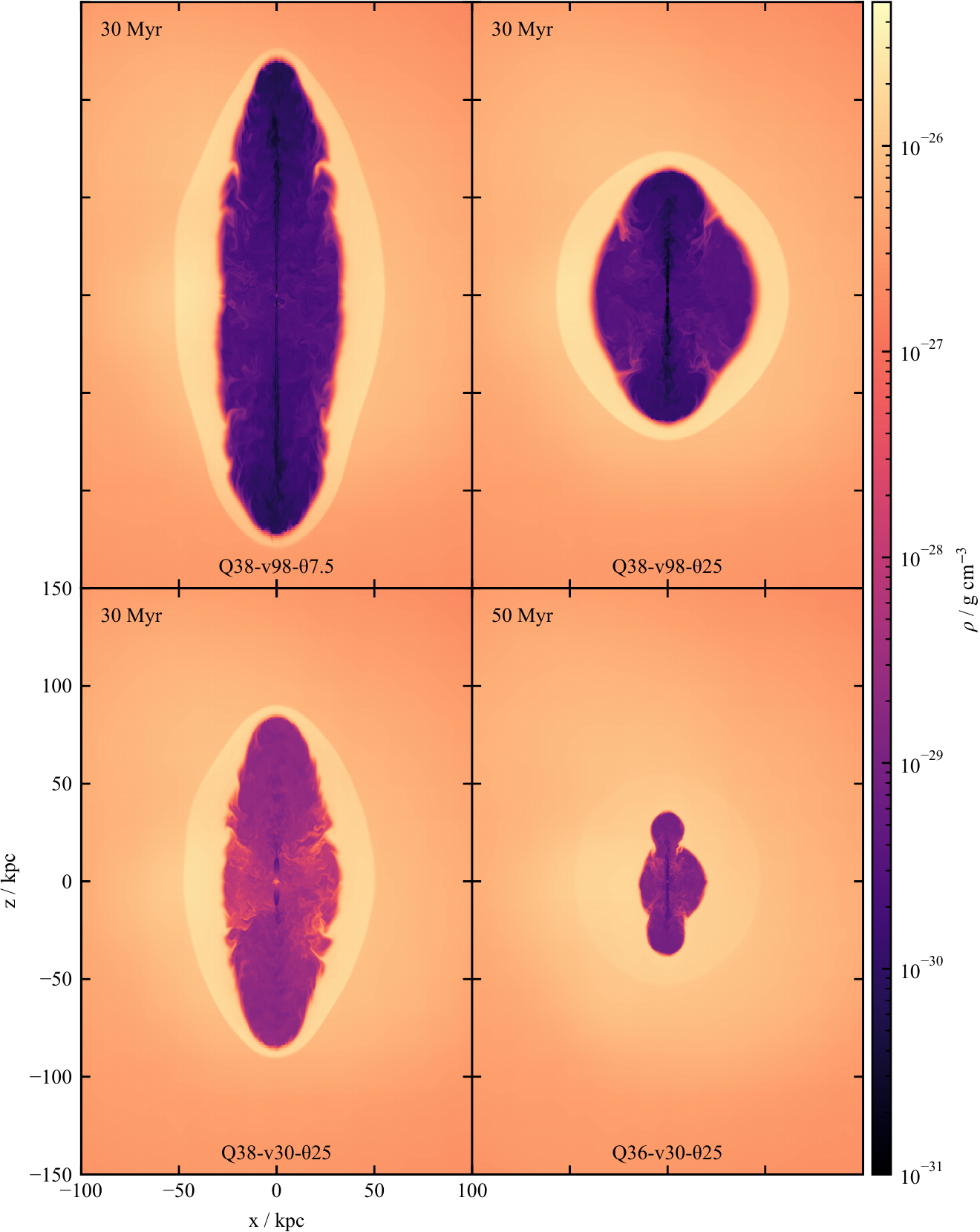}
      \caption{
        Midplane density slices in the $y$-axis of the four high power simulations.
        The simulation label is given at the bottom of each panel, while
        the time at which the density slice is made is in the top-left corner
        of each panel.
      }
      \label{fig:cd_jet_population_density}
    \end{figure*}

    \begin{figure*}
      \centering
      \includegraphics{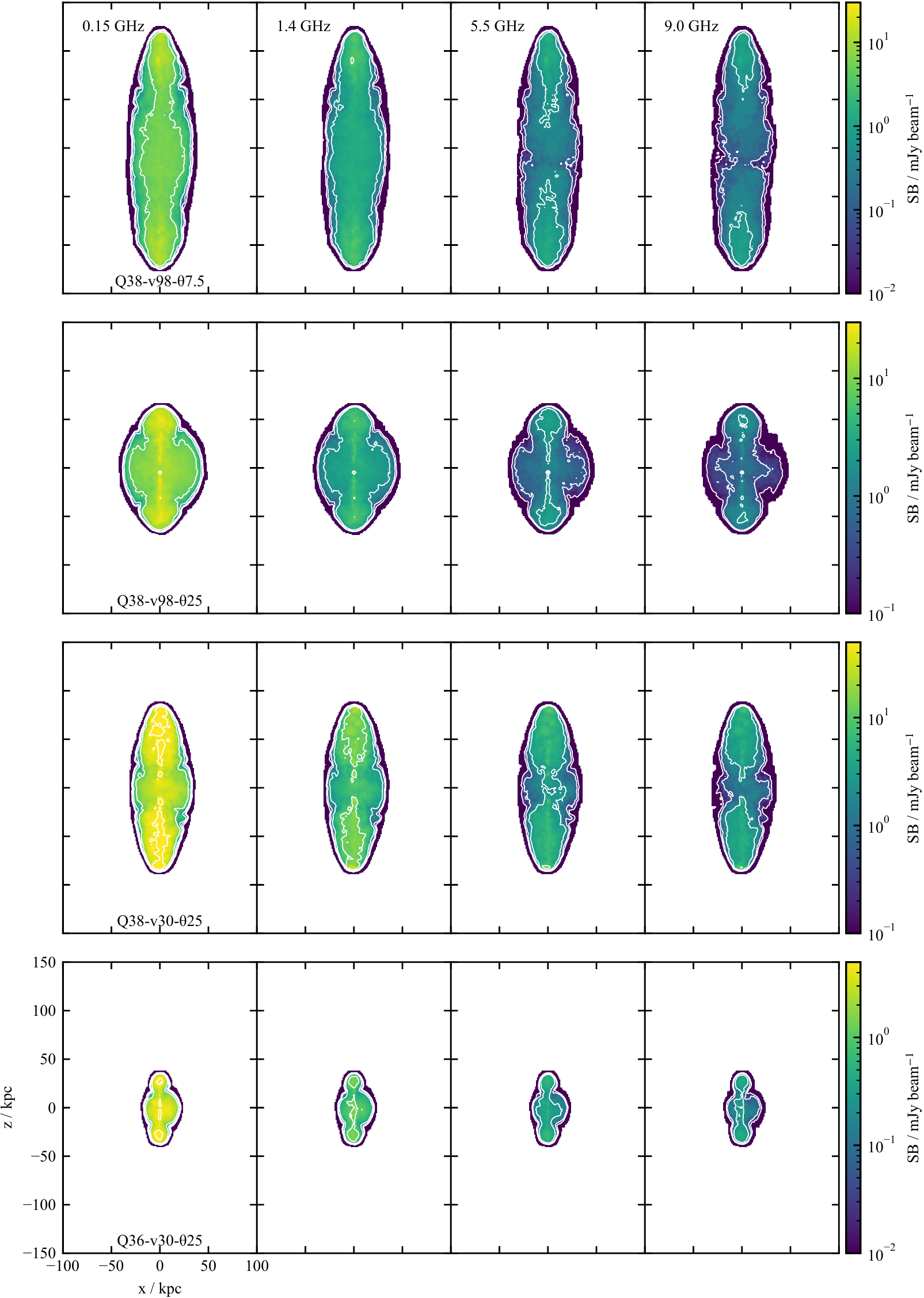}
      \caption{
        Synthetic surface brightness maps of the four high power simulations, at $0.15, 1.4, 5.5$, and $9.0\,\textrm{GHz}$ (from left to right).
        Simulation times are as in \cref{fig:cd_jet_population_density}.
        Individual simulation surface brightness limits are chosen to highlight source structure, and are constant for a given simulation across all frequencies.
        There are five contours evenly spaced in log-space between the surface brightness limits.
        The sources are observed in the plane of the sky with a $1.5\,\textrm{arcsec}$ FWHM Gaussian beam.
      }
      \label{fig:cd_jet_population_surface_brightness}
    \end{figure*}

    \begin{figure}
      \includegraphics{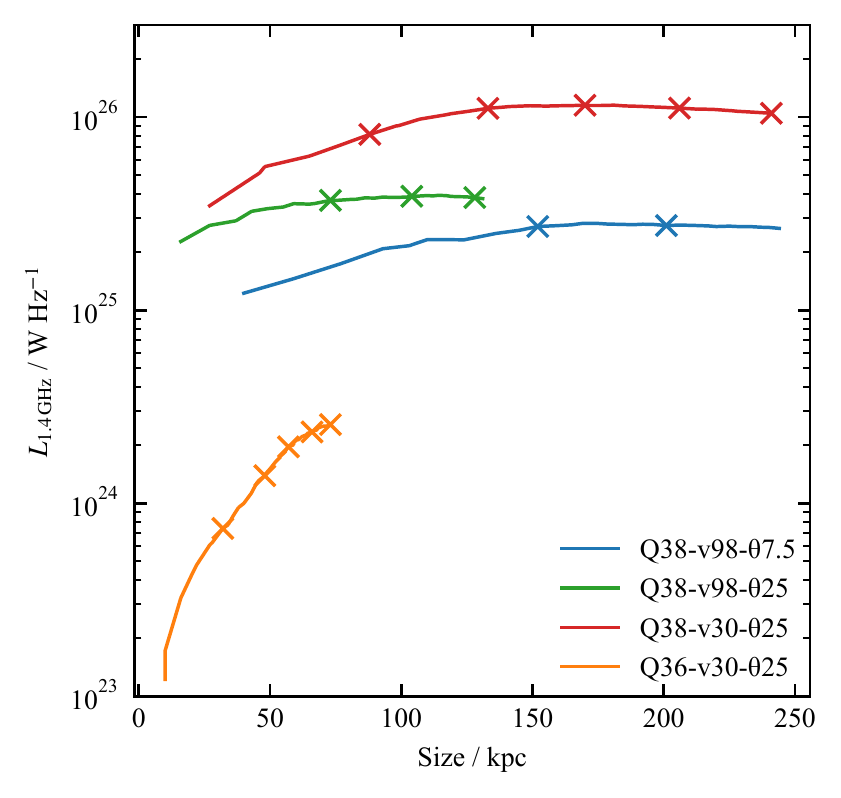}
      \caption{
        Total source size vs.~ $1.4\,\textrm{GHz}$ luminosity evolution with time, for the high power simulations.
        The lobe size is measured as the distance from the injection point to the most distant point of emission 2 dex below the maximum surface brightness.
        Crosses are placed in $10\myr$ increments for all simulations.
      }
      \label{fig:cd_frII_pd_tracks}
    \end{figure}

    \begin{figure*}
      \centering
      \includegraphics{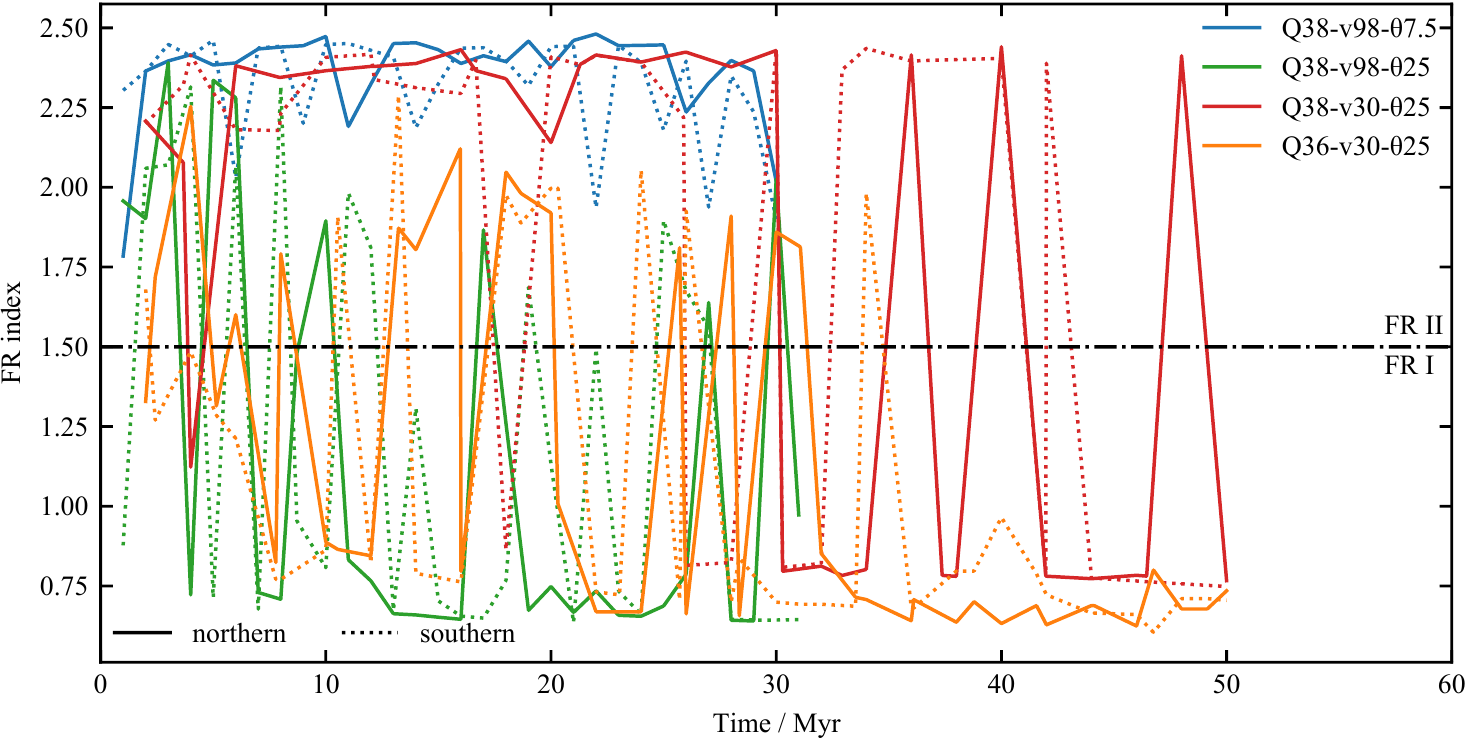}
      \caption{
        Fanaroff-Riley index as a function of time for individual lobes of the high power simulations.
        For all simulations the FR index for the northern lobe is plotted as the solid lines, while the southern lobe is plotted as the dotted lines.
      }
      \label{fig:cd_frII_fr_index}
    \end{figure*}

    \begin{figure*}
      \centering
      \includegraphics{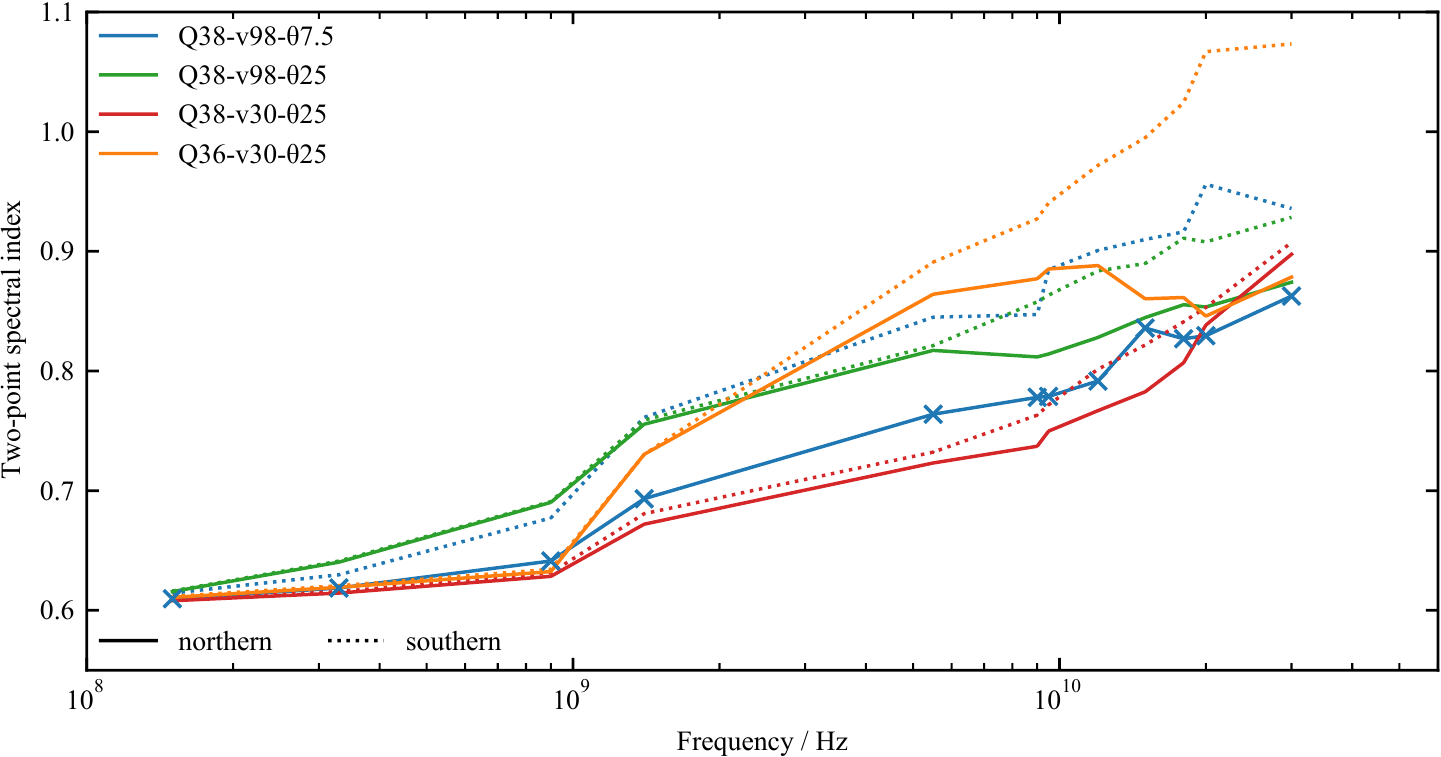}
      \caption{
        Integrated lobe spectral index for the high power simulations.
        The crosses mark the midpoints of the two-frequency differences used.
        Line styles are as in \cref{fig:cd_frII_fr_index}.
      }
      \label{fig:cd_frII_total_spectral_index}
    \end{figure*}

    In this section, we present the first four simulations listed in \cref{tbl:cd_jet_population_parameters}, covering two jet powers ($Q=10^{36},10^{38}\,\textrm{W}$), two jet half-opening angles ($\theta_\textrm{j}=7.5\degr,25\degr$), and two jet velocities ($v_\textrm{j}=0.3c, 0.98c$).
    We begin by considering the dynamics and morphology of the jets in these simulations in \cref{sec:cd_frII_dynamics_and_morphology}, and then discuss their observable radio signatures in \cref{sec:cd_frII_radio_observables}.
    
    \subsubsection{Dynamics and morphology}
    \label{sec:cd_frII_dynamics_and_morphology}

      In \cref{fig:cd_jet_population_density} we show midplane density slices of four CosmoDRAGoN simulations of faster jets, which we expect to produce FR II morphology based on the jet velocities as detailed in \cref{tbl:cd_jet_population_parameters}.
      The three high power jets are shown at $t=30\myr$, while the low power jet is shown at $t=50\myr$.

      There are significant differences in the cocoon and bow shock structure of the four simulations. 
      The high power, strongly relativistic, jets (Q38-v98-$\uptheta$7.5, Q38-v98-$\uptheta$25) inflate wide, low-density cocoons.
      The contact discontinuity between the cocoon and shocked material is smooth, with little evidence of turbulent mixing.
      For both half-opening angles, the initially conical flow quickly collimates into a low-density jet beam.
      In both cases, this beam is disrupted before the terminal shock; however, it remains coherent for longer distances in the narrow opening-angle simulation, due to a narrower collimated jet width and higher collimated density.
      This causes the elongated morphology in Q38-v98-$\uptheta$7.5, where the more collimated jet distributes its thrust over a smaller area.
      Both the strongly relativistic simulations exhibit the FR II characteristics of a collimated jet, low-density cocoon, and well-defined jet heads.

      The slower jets (Q38-v30-$\uptheta$25, Q36-v30-$\uptheta$25) produce many of the same characteristics as their faster counterparts.
      A low-density cocoon is formed (albeit with a higher density than in the relativistic case), and a bow shock is formed.
      We note the large difference in volume between the (weakly) shocked and cocoon material in the Q36-v30-$\uptheta$25 simulation.
      The radio-emitting electrons occupy a very different volume to the bow-shock, being generally confined to the cocoon.
      Hence, the morphology of the observable radio source and the feedback it produces are likely to be very different; a detailed investigation of this point will be presented in a future paper.
      Both slow jets undergo an initial collimation event and, in the case of the lower power jet, remain collimated for $\sim30\kpc$ at $t=50\myr$ before transitioning to turbulent flow.
      Meanwhile, simulation Q38-v30-$\uptheta$25 exhibits the shock morphology found in simulations of FR Is downstream of recollimation reminiscent of flaring points observed in FR I radio galaxies \citep{Krause2012}.
      This produces a morphology that resembles a lobed FR I, with no clearly defined jet head or terminal shock in the jet head region.

    \subsubsection{Observable signatures}
    \label{sec:cd_frII_radio_observables}

      Low redshift ($z=0.05$) synthetic radio surface brightness maps at four frequencies are shown for all four simulations in \cref{fig:cd_jet_population_surface_brightness}.
      These maps are created using the PRAiSE method for calculating spectral aging from hydrodynamic simulations \citep[see Sect 2.2,][]{YatesJonesEA22}, and model both adiabatic and radiative loss processes for synchrotron-emitting electrons.
      We use a pressure threshold of $\epsilon_p = 5$, corresponding to a minimum Mach number of $\mathcal{M} \sim 2.24$, tracking particle acceleration only at strong shocks.
      As magnetic fields are not included in the simulations presented here, a constant departure from equipartition of magnetic ($U_\textrm{B}$) and particle ($U_\textrm{e}$) energy densities is assumed, where we take the hydrodynamic pressure in the simulation equal to the pressure in radio-emitting leptons with an equipartition factor $\eta = U_\textrm{B}/U_\textrm{e} = 0.03$.
      This value is representative of moderate power radio sources \citep{Croston2014,Ineson2017}.
      An electron spectral index of $s=2.2$ and power-law electron energy distribution with $\gamma_\textrm{min}=500,  \gamma_\textrm{max}=10^{5}$ are used; these parameters are typical of FR II radio sources \citep[see][and references therein]{YatesJonesEA21}.
      An observing beam with a full width half maximum (FWHM) of $1.5\,\textrm{arcsec}$ is used, and relativistic beaming effects are included.

      Classic double radio lobes are produced for all simulations.
      When viewed in the plane of the sky, the two high power, fast jet simulations (Q38-v98-$\uptheta$7.5, Q38-v98-$\uptheta$25) produce clear hotspots and edge-brightened lobes associated with FR IIs for at least part of the simulation time.
      For the high power, slow jet (Q38-v30-$\uptheta$25), well-defined radio lobes are produced, and faint hotspots are observed despite the lack of a clear jet head and terminal shock in the underlying morphology.
      We propose that the hotspots present in these radio sources are indicative of forward flowing electrons shocked near the flaring point, and that a hotspot is formed is due to a higher concentration of emitting material, rather than the presence of strong shocks.
      Meanwhile, the low power, slow jet has faint hotspots, which are of similar brightness to emission along the jet.
      This leads to a more FR I-like morphology, which is expected to change with viewing angle.
      We discuss the dependence of observed morphology on viewing angle in \cref{sec:cd_radio_observables}.

      In \cref{fig:cd_frII_pd_tracks} we plot size-luminosity (PD) tracks for all four high power simulations.
      The total luminosity is found by integrating the surface brightness in the edge-on orientation for each timestep, while the source size is measured from the surface brightness map as the maximum distance from the core that has a surface brightness within two orders of magnitude of the brightest pixel.
      We find that all simulations have different  tracks through the size-luminosity diagram, indicating the importance of environment and jet parameters (speed and half-opening angle) on total source luminosity.
      The jet power is not the only factor; the total luminosity depends on cocoon volume (an increased volume leads to both a larger emitting volume and increased adiabatic losses) and the total source length is dependent on jet recollimation (as shown in \cref{fig:cd_jet_population_density}).

      The Fanaroff-Riley (FR) index is a useful tool for classifying observed radio source morphology.
      Following \citet{Krause2012}, the FR index for each lobe is calculated as $\textrm{FR}=2 x_\textrm{bright} / x_\textrm{length} + 1/2$ at $150\,\textrm{MHz}$, where for each lobe $x_\textrm{bright}$ is the radius at the brightest point, and $x_\textrm{length}$ is the lobe length.
      This produces indices of $0.5 < \textrm{FR} < 1.5$ and $1.5 < \textrm{FR} < 2.5$ for radio sources with FR I and FR II morphology respectively.
      In \cref{fig:cd_frII_fr_index} we show the evolution of FR index with time for these first four simulations; the northern and southern lobes are plotted separately as the solid and dotted lines respectively.
      We find that the Q38-v98-$\uptheta$7.5 and Q38-v30-$\uptheta$25 simulations are consistently within the FR II range for the first $\sim30\myr$, indicating clear hotspots.
      For Q38-v98-$\uptheta$25, the FR index evolution is very noisy.
      This simulation is expected to start out as an FR II before transitioning to FR I morphology as pressure equilibrium is reached \citep{Krause2012}.
      This transition is gradual as the cluster weather causes pressure fluctuations near the tip of the lobe. 
      A similar effect is seen for the Q36-v30-$\uptheta$25 simulation; at later times, however, the FR index largely indicates FR I morphology.
      The FR index is too stochastic to reliably comment on differences between the northern and southern lobes.

      Finally, in \cref{fig:cd_frII_total_spectral_index} we plot the spectral index over the frequency range $10^8-10^{10}\,\textrm{Hz}$ for the same simulation times as \cref{fig:cd_jet_population_surface_brightness}; as with the FR index, the spectral index for the northern lobe is plotted as the solid line, while for the southern lobe it is plotted as the dotted line.
      If no losses were included, and the only emission was due to recently accelerated electrons, the integrated spectral index would be approximately constant, indicating a power-law in electron energies with a constant slope.
      The inclusion of radiative losses steepens the spectrum at higher frequencies, in all simulations.
      Differences in spectral index between the northern and southern lobes (e.g. in Q36-v30-$\uptheta25$) are observed, even in this relaxed cluster.
      This is due to the environment affecting the morphology and dynamics of each lobe differently, leading to different electron energy distributions in the two lobes.

  \subsection{Low power, slow radio jets in cosmological environments}
  \label{sec:cd_frI}

    \begin{figure*}
      \centering
      \includegraphics{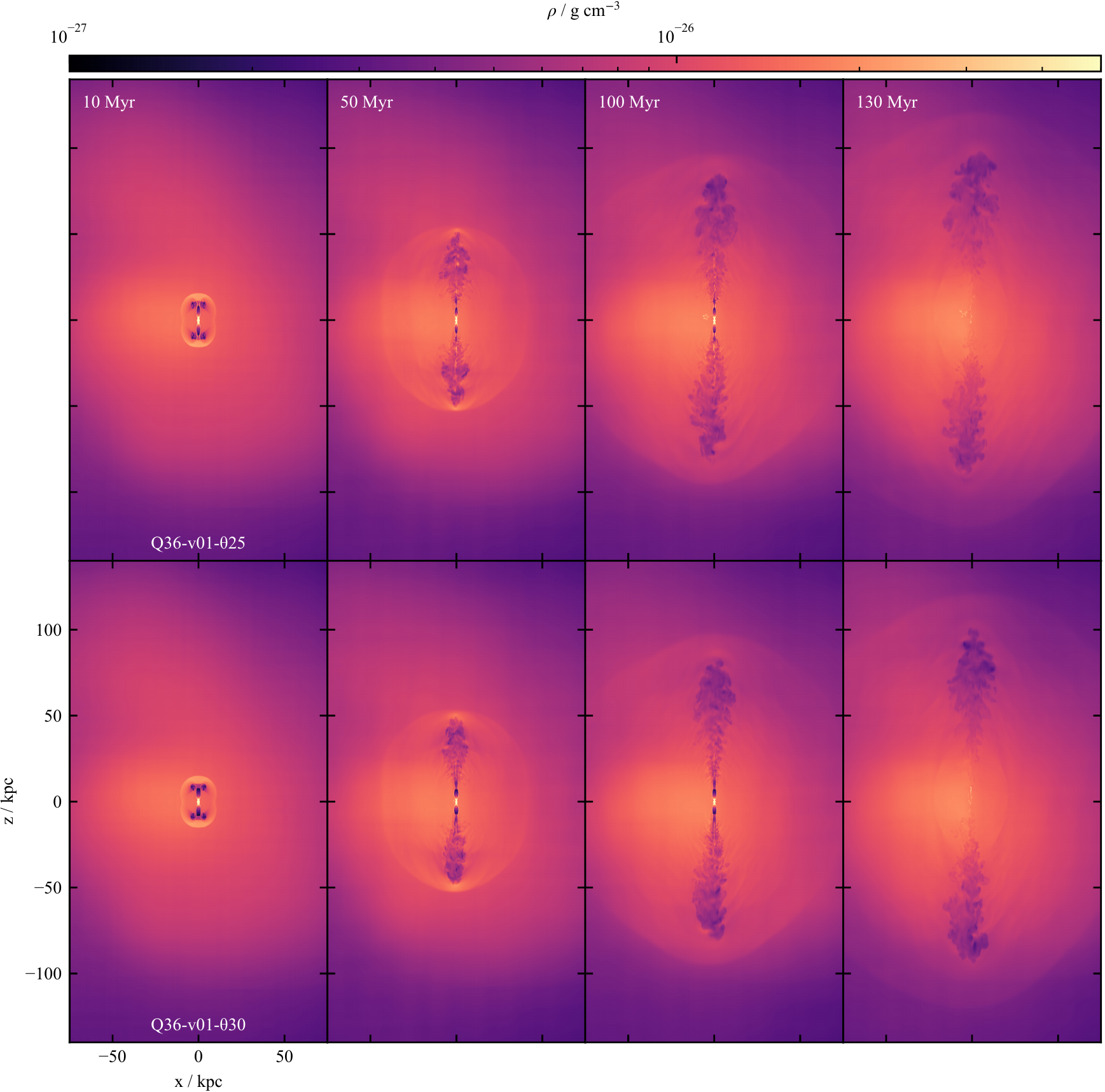}
      \caption{
        Midplane density slices in the y-axis of the simulations reproducing FR I morphological features, at times $t=10,50,100,130\myr$, for Q36-v01-$\uptheta$25 (top) and Q36-v01-$\uptheta$30 (bottom).
      }
      \label{fig:cd_frI_density_evolution}
    \end{figure*}

    \begin{figure*}
      \centering
      \includegraphics{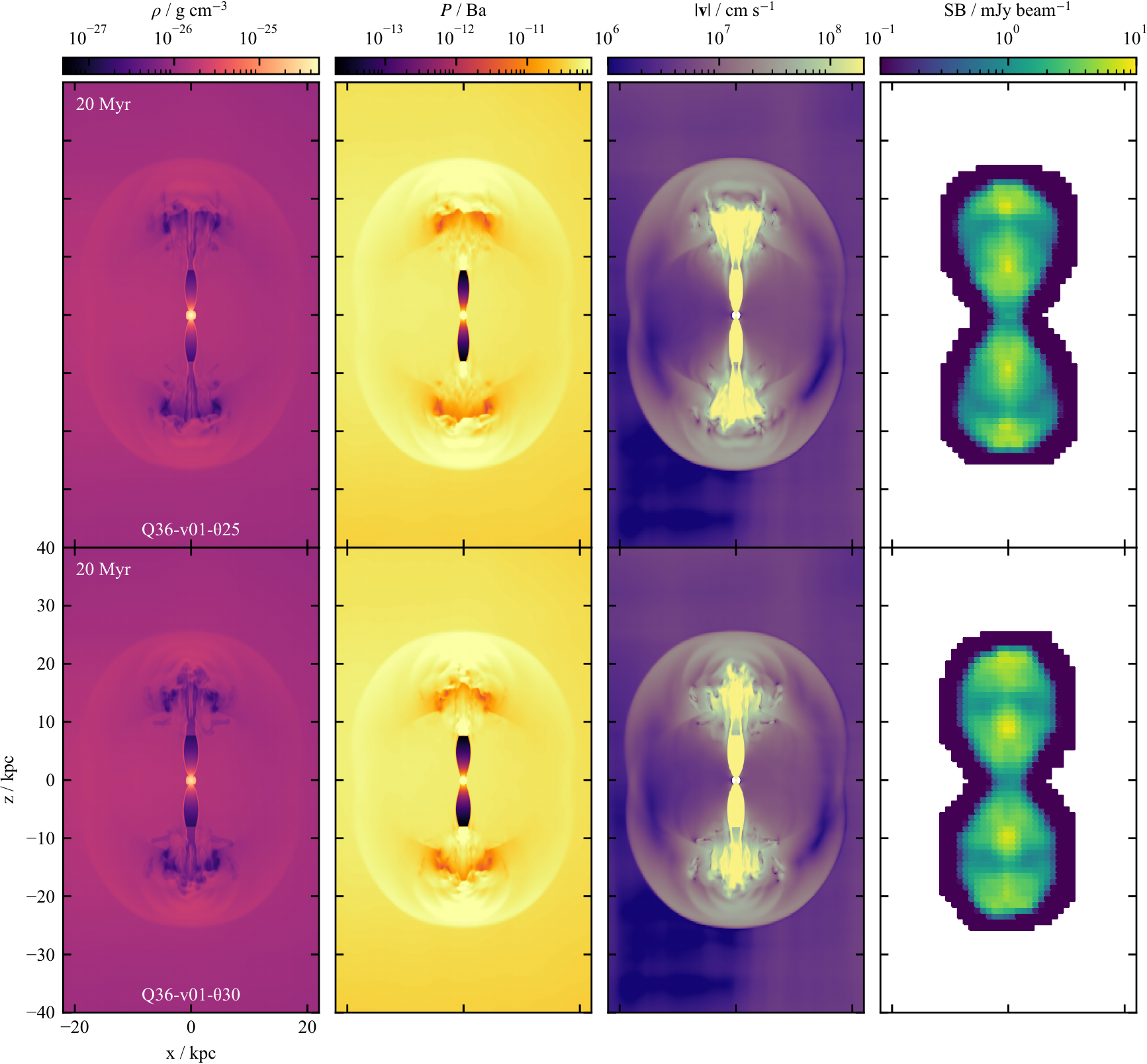}
      \caption{
        The recollimation and flaring region at time $t=20\myr$ for the low power, slow jet simulations.
        From left to right, the quantities plotted are: density, pressure, total velocity, and $1.4\,\textrm{GHz}$ surface brightness.
        The first three columns are midplane slices of the quantity through the y-axis.
        The observing properties for the final column are as described in \cref{sec:cd_radio_observables}, and the radio source is in the plane of the sky, oriented to match the hydrodynamic slices.
        Rows are as in \cref{fig:cd_frI_density_evolution}.
      }
      \label{fig:cd_frI_zoomed}
    \end{figure*}

    \begin{figure}
      \includegraphics{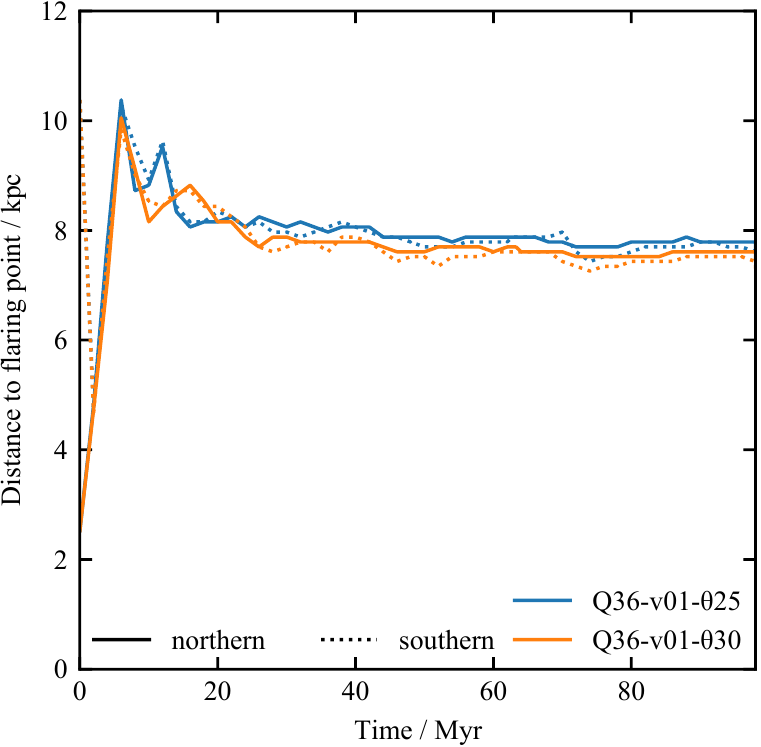}
      \caption{
        Flaring region distance as a function of time for the low power, slow jet simulations.
        The distance for the northern lobe is plotted as the solid line, while the distance for the southern lobe is plotted as the dotted line.
      }
      \label{fig:cd_frI_flaring_region}
    \end{figure}

    \begin{figure*}
      \includegraphics{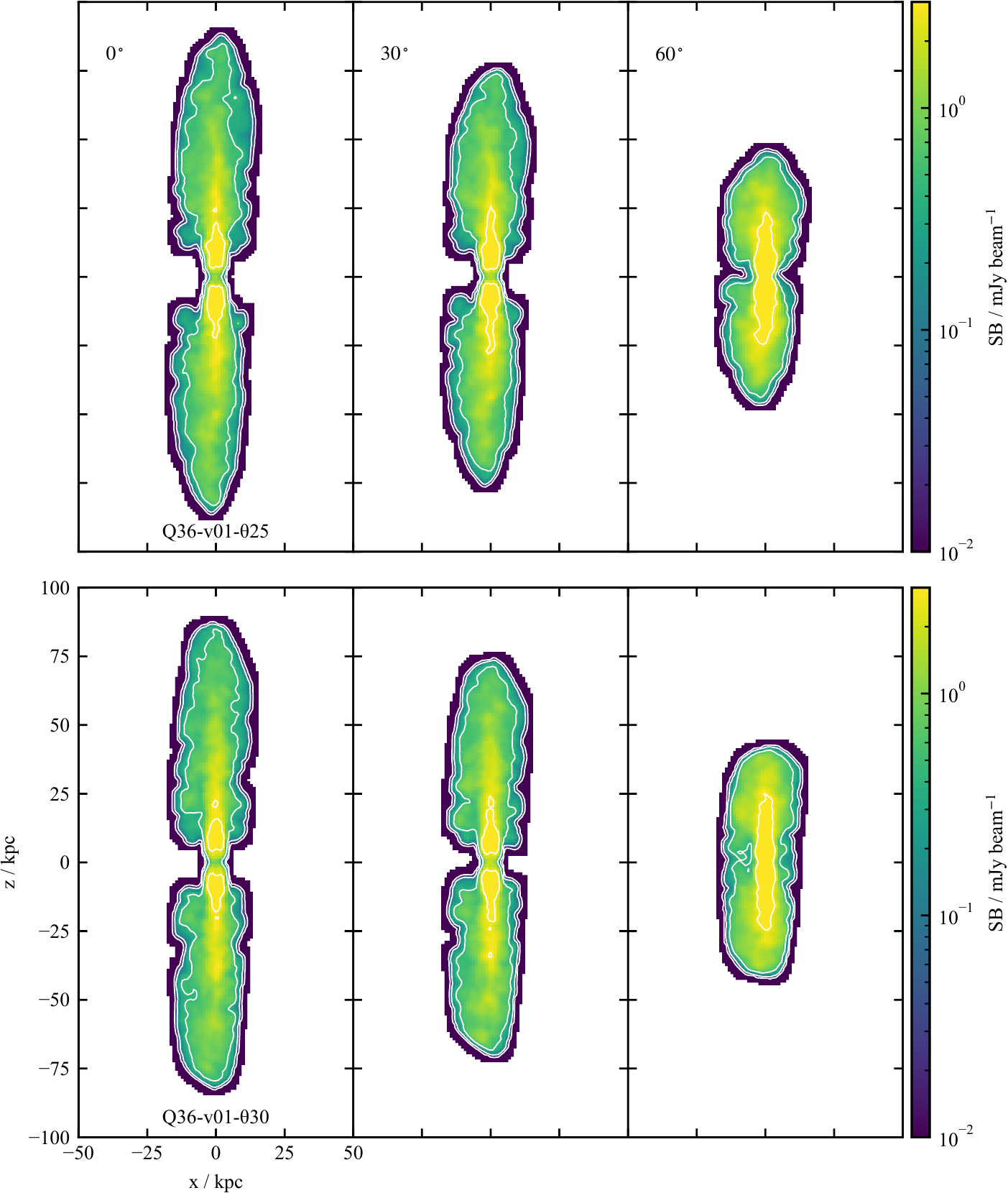}
      \centering
      \caption{
        Synthetic surface brightness maps of the low power, slow jet simulations, at the observing frequency $1.4\,\textrm{GHz}$ and time $t=100\myr$.
        Three different source orientations are shown: plane of the sky, or inclined $30\degr$ or $60\degr$ with respect to the observer.
        As in \cref{fig:cd_jet_population_surface_brightness}, the sources are observed with a $1.5\,\textrm{arcsec}$ FWHM Gaussian beam.
        The contours are at $0.01, 0.07, 0.45, 3.0 \,\times 1\,\textrm{mJy beam}^{-1}$.
        Rows are as in \cref{fig:cd_frI_density_evolution}.
      }
      \label{fig:cd_frI_surface_brightness}
    \end{figure*}

    \begin{figure}
      \includegraphics{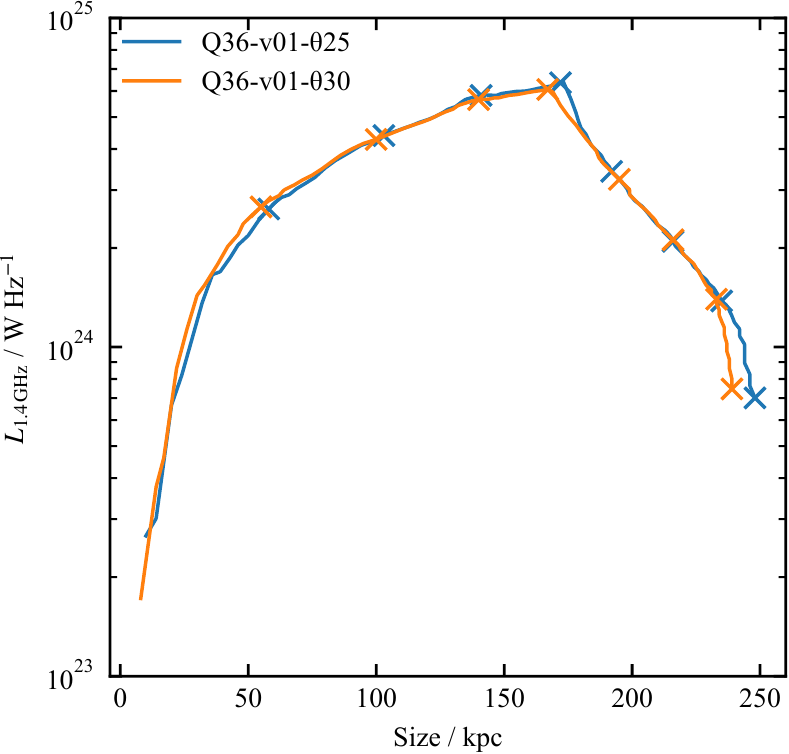}
      \caption{
        Total source size vs.~ $1.4\,\textrm{GHz}$ luminosity evolution with time, for the low power, slow jet simulations.
        The lobe size is measured as the distance from the injection point to the furthest point of emission 2 dex below the maximum.
        Crosses are placed in $25\myr$ increments for both simulations.
      }
      \label{fig:cd_frI_pd_tracks}
    \end{figure}

    \begin{figure}[t]
      \includegraphics{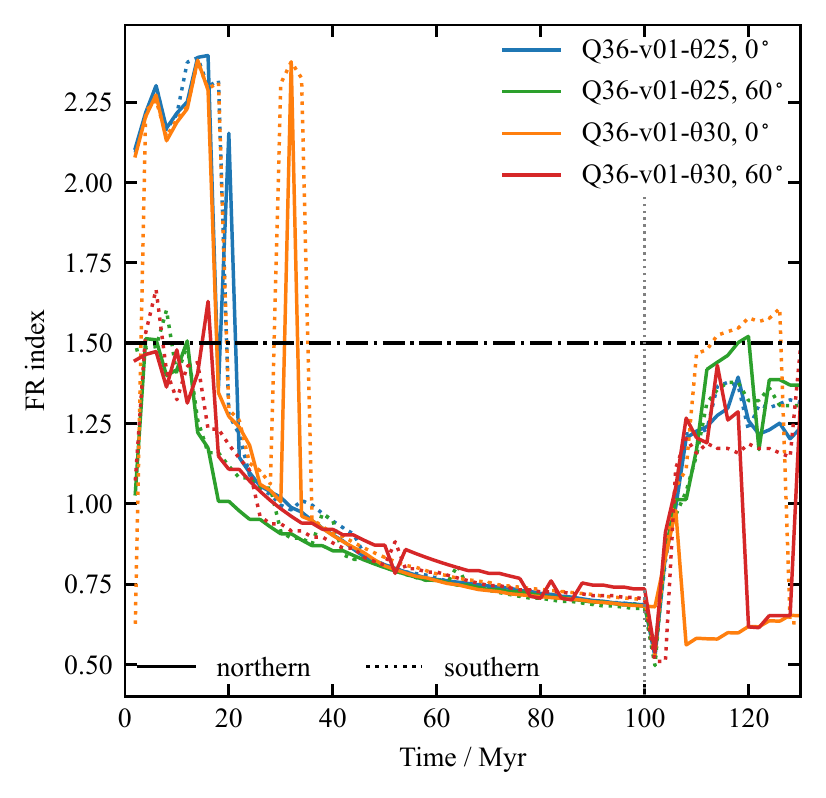}
      \caption{
        Fanaroff-Riley (FR) index as a function of time for individual lobes in the low power, slow jet simulations.
        Line styles are as in \cref{fig:cd_frI_flaring_region}.
        A vertical dotted line is drawn at $t=100\myr$, when the jets switch off.
      }
      \label{fig:cd_frI_fr_index}
    \end{figure}

    \begin{figure}
      \includegraphics{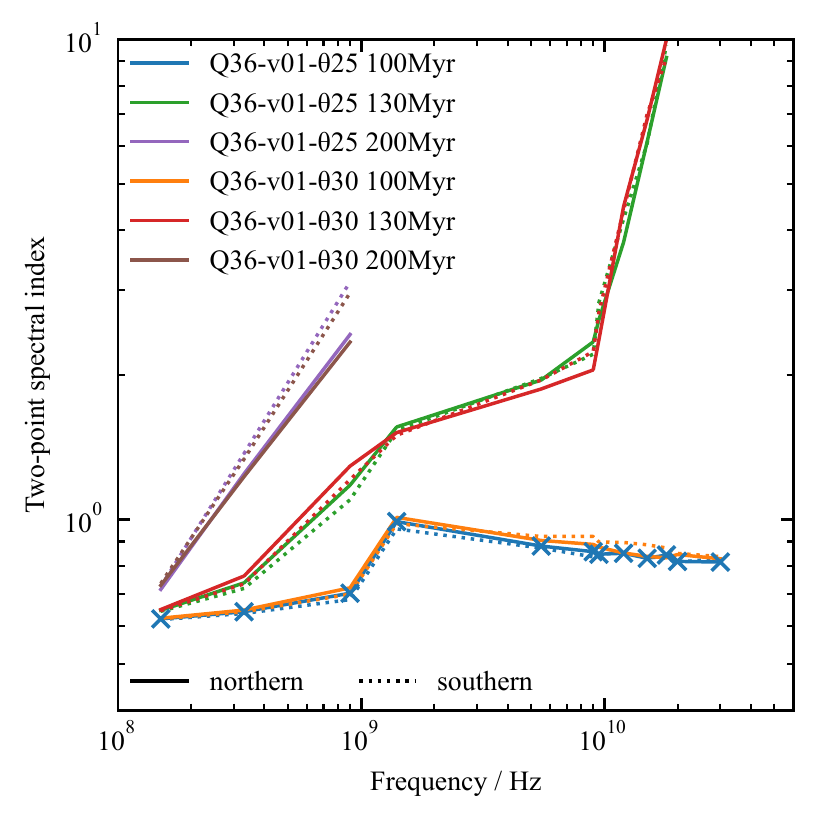}
      \caption{
        Integrated lobe spectral index for the low power, slow jet simulations, at $t=100, 130, 200\myr$.
        The crosses mark the midpoints of the two-frequency differences used.
        Line styles are as in \cref{fig:cd_frI_flaring_region}.
      }
      \label{fig:cd_frI_total_spectral_index}
    \end{figure}

    \begin{figure}[t]
      \includegraphics{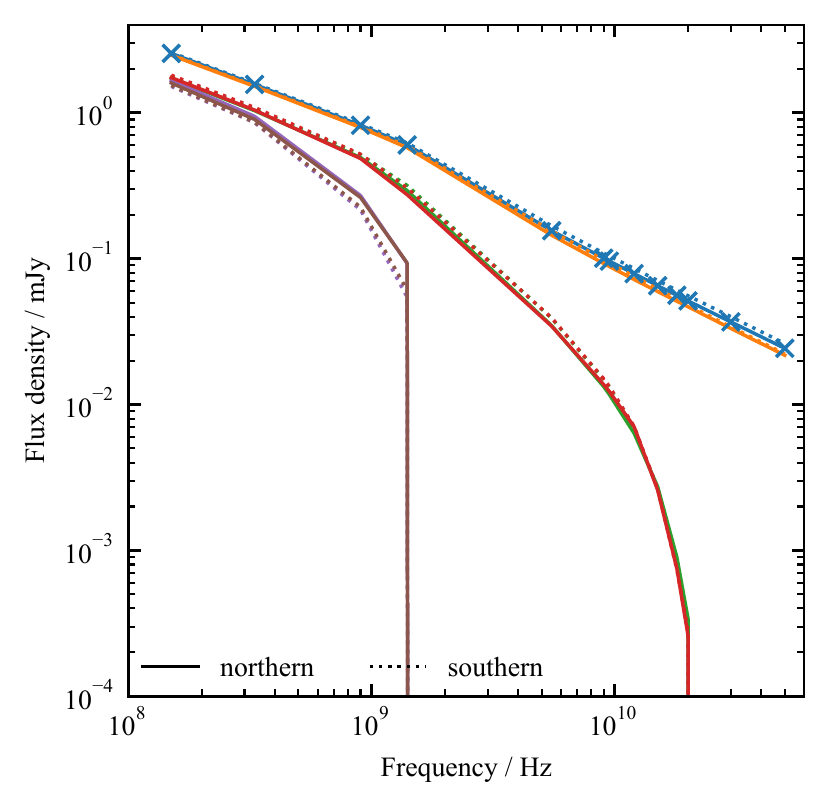}
      \caption{
        Integrated lobe spectra for the low power, slow jet simulations, at $t=100, 130, 200\myr$.
        The crosses mark the frequencies at which the integrated flux density is calculated.
        Line styles and colours are as in \cref{fig:cd_frI_total_spectral_index}.
      }
      \label{fig:cd_frI_spectra}
    \end{figure}

    \begin{figure*}
      \centering
      \includegraphics{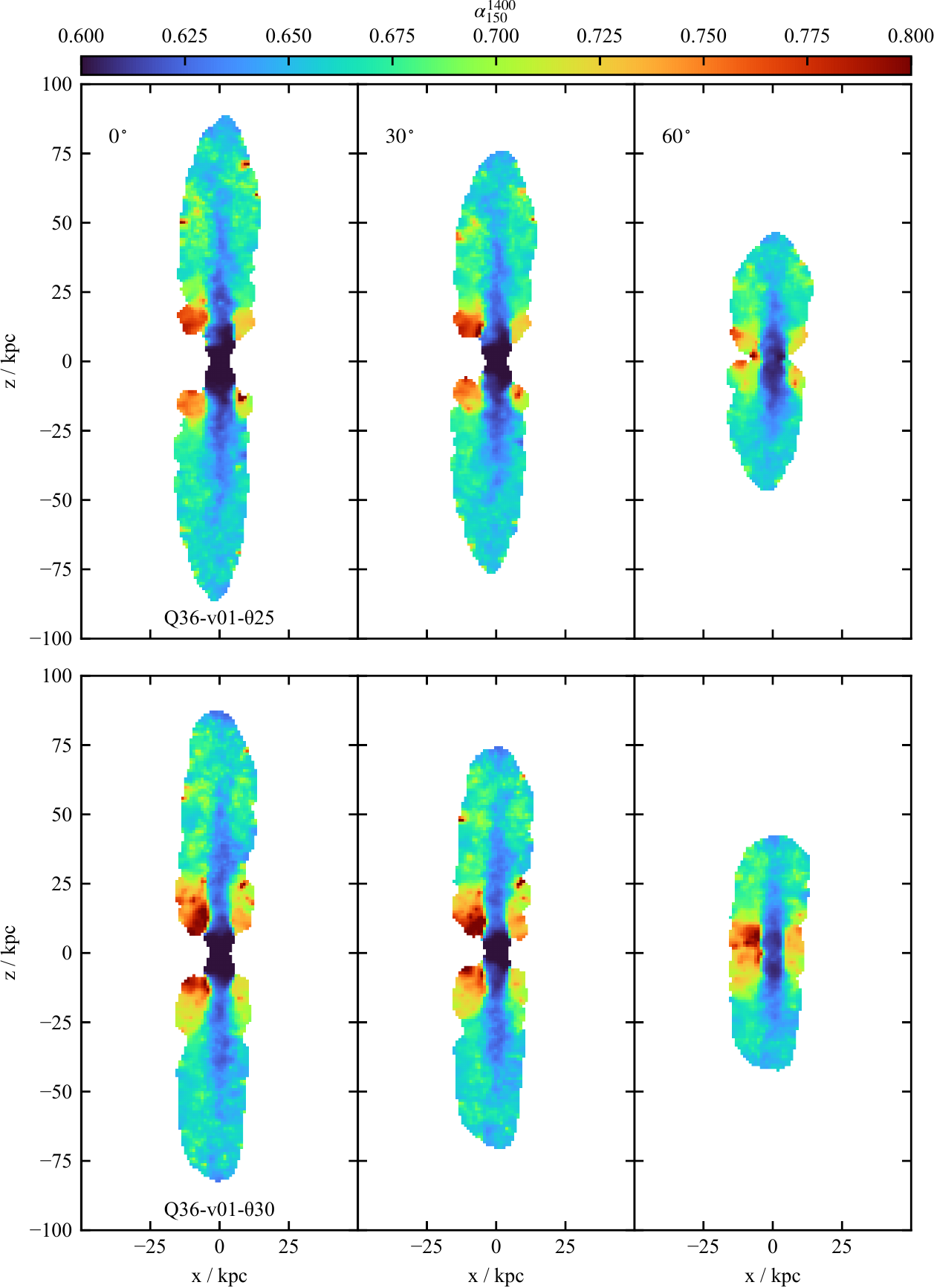}
      \caption{
        Synthetic spectral index maps of the low power, slow jet simulations at time $t=100\myr$ and $z=0.05$, between $\nu_\textrm{high}=1.4\,\textrm{GHz}$ and $\nu_\textrm{low}=150\,\textrm{MHz}$.
        The source orientations and rows are as in \cref{fig:cd_frI_surface_brightness}.
      }
      \label{fig:cd_frI_spectral_index_map_low}
    \end{figure*}

    \begin{figure*}
      \centering
      \includegraphics{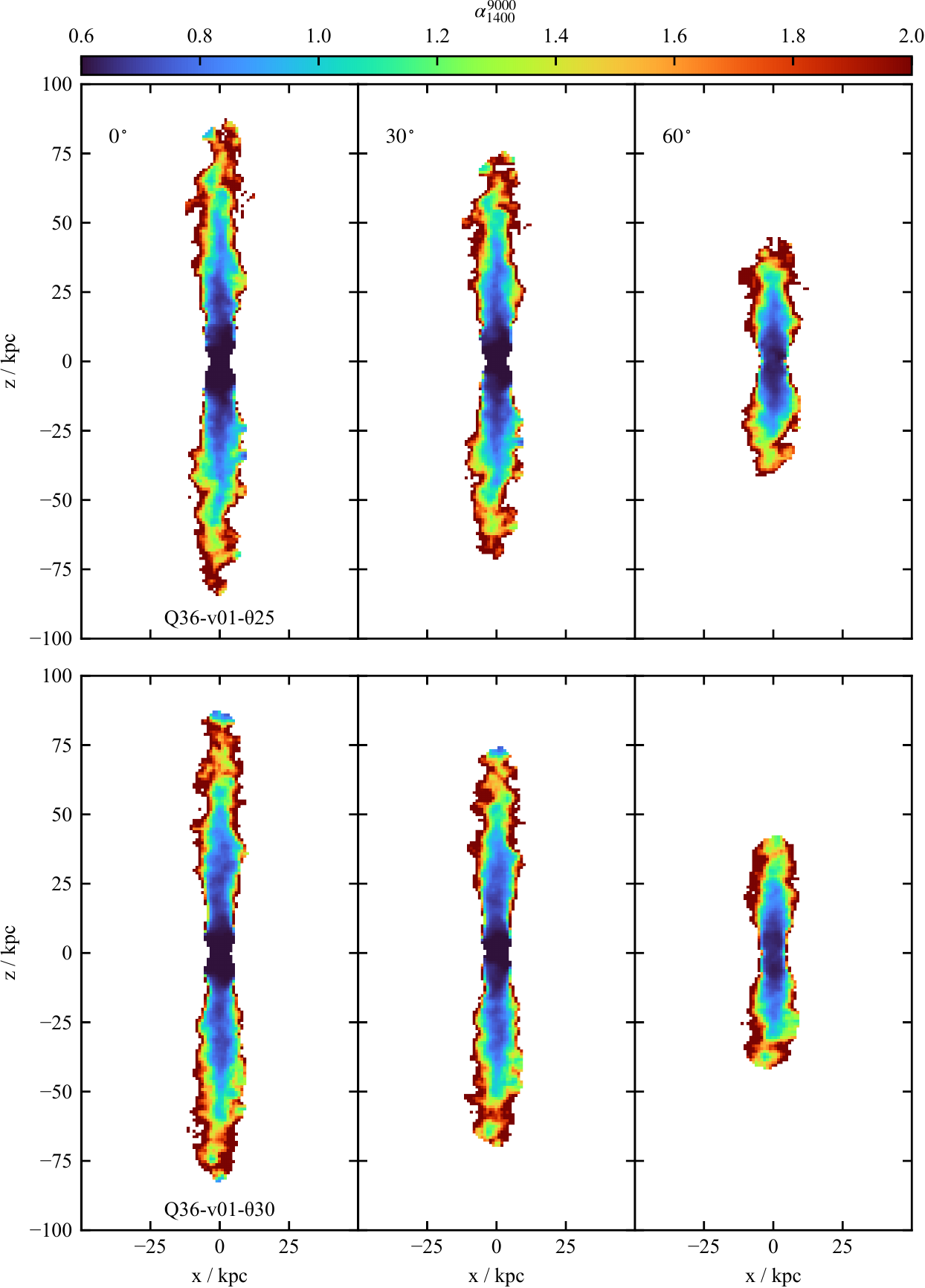}
      \caption{
        Synthetic spectral index maps of the low power, slow jet simulations at time $t=100\myr$, between $\nu_\textrm{high}=9.0\,\textrm{GHz}$ and $\nu_\textrm{low}=1.4\,\textrm{GHz}$.
        The source orientations and rows are as in \cref{fig:cd_frI_surface_brightness}.
      }
      \label{fig:cd_frI_spectral_index_map_high}
    \end{figure*}

    \begin{figure*}
      \centering
      \includegraphics{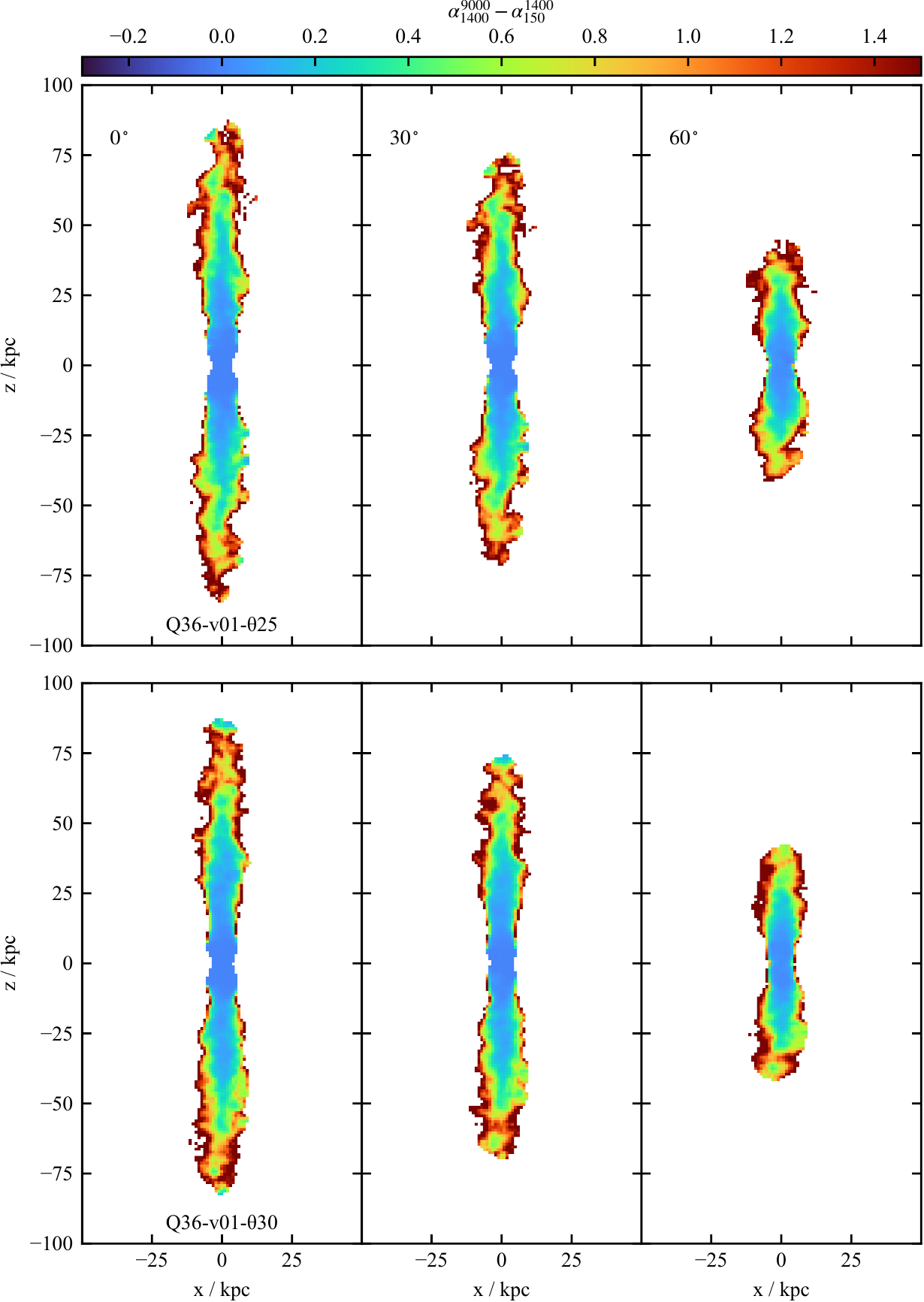}
      \caption{
        Synthetic spectral curvature maps of the low power, slow jet simulations at time $t=100\myr$, $\alpha^{9000}_{1400} - \alpha^{1400}_{150}$.
        The source orientations and rows are as in \cref{fig:cd_frI_surface_brightness}.
      }
      \label{fig:cd_frI_spectral_curvature_map}
    \end{figure*}

    \begin{figure*}
      \centering
      \includegraphics{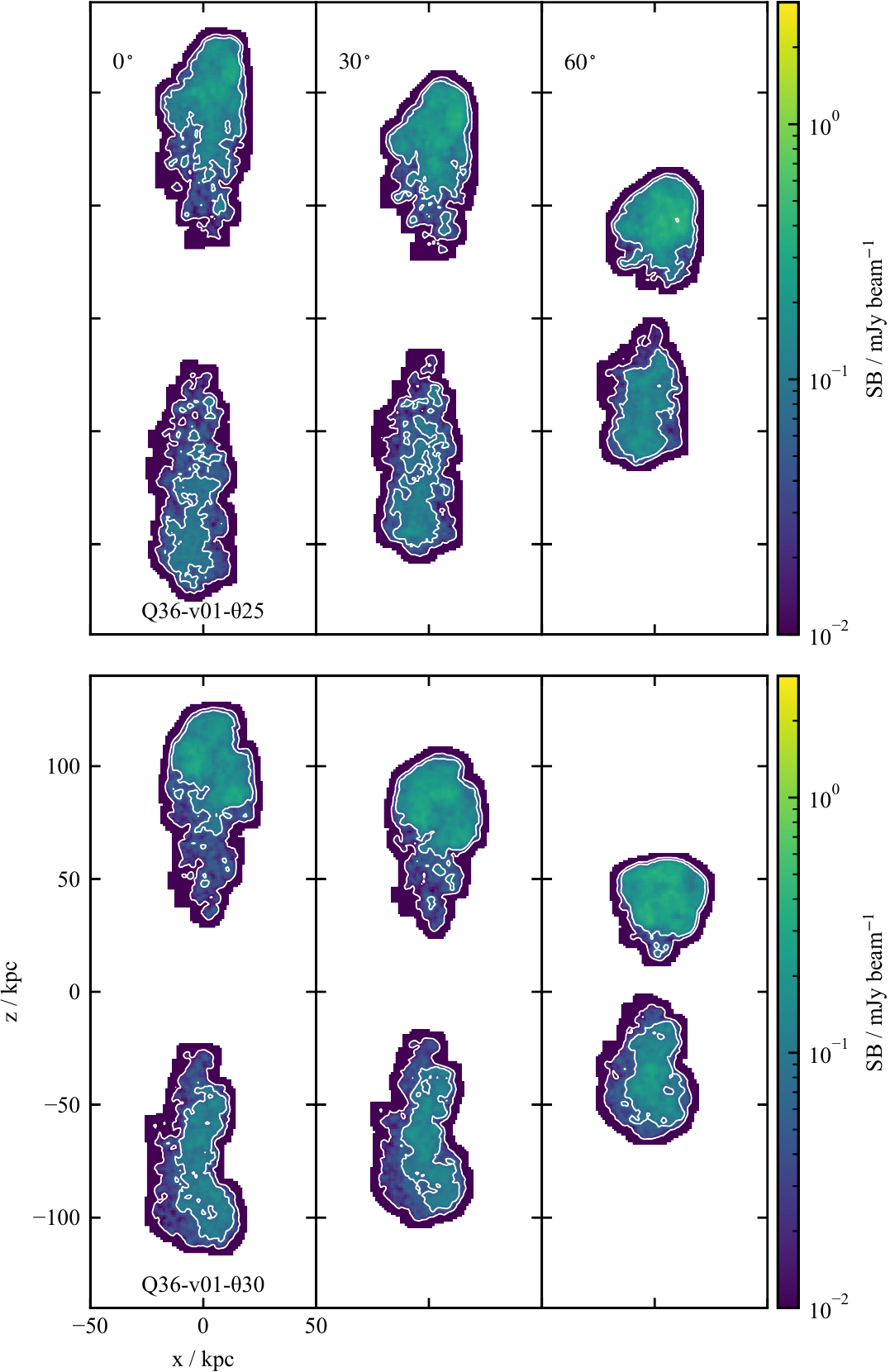}
      \caption{
        Synthetic surface brightness maps of the low power, slow jet simulations as in \cref{fig:cd_frI_surface_brightness}, at time $t=200\myr$.
        The radio source has been switched off for $100\myr$.
      }
      \label{fig:cd_frI_surface_brightness_remnant}
    \end{figure*}

    In this section, we focus on the large-scale evolution of low power, slow jets.
    The full simulation suite will include several jets with FR I-like morphology in a variety of environments; here, however, we limit our focus to two specific simulations with very large half-opening angles.

    The two jets are launched into environment \textit{002-0003}, with a velocity $v_\textrm{j}=0.01c$ and power $Q=10^{36}\,\textrm{W}$.
    Each jet is active for $100\myr$, after which the jet is switched off and the remnant evolution is followed.
    The half-opening angle is different for both jets; simulation Q36-v01-$\uptheta$25 has a half-opening angle of $25\degr$ (on the cusp of the FR I / FR II transition for conical jets, see \citet{Krause2012}), while simulation Q36-v01-$\uptheta$30 has a half-opening angle of $30\degr$, placing it well into the regime where an FR I morphology should be produced.
    Our simulated jets have lower velocities than the initial velocities of observed FR-I jets \citep{Laing2014}; however, these are representative of both heavily mass-loaded jets \citep{Perucho2014} and massive, slow AGN outflows in cosmological simulations.

    \subsubsection{Dynamics and morphology}
    \label{sec:cd_frI_dynamics_morphology}
    
      The evolution of density for the two low power, slow jet simulations are shown in \cref{fig:cd_frI_density_evolution}.
      The overall morphology of the two simulations is similar: both produce a bow shock that propagates through the cluster, and exhibit the low-density plume morphology associated with FR Is at later times.
      Both jets expand conically with laminar flow, before reaching a flaring point and transitioning to turbulence.
      This reproduces some of the characteristics identified by \citet{Laing2014} for the transition region.
      There is a clear site of particle acceleration and Kelvin-Helmholtz instabilities evolving into turbulence are also present.
      While other dynamical effects including centrifugal and Rayleigh-Taylor instabilities \citep{Gourgouliatos2018,Matsumoto2017} or mixing at the jet boundary due to jet-star interactions \citep{Perucho2020} may also contribute, our simulations show that shearing plays a key role in jet evolution.
      The flow decelerates after the flaring point, where the particle acceleration occurs.
      In later snapshots, the widening after the flaring point can be observed.
      After $100\myr$, the jet is switched off, with the lobes entering a remnant phase.
      At this point, the effects of a dynamic environment are more prominent, as the lobes begin to rise buoyantly.
      

      In \cref{fig:cd_frI_zoomed} the flaring and recollimation region is shown at $t=20\myr$.
      The flaring point is identified as a discontinuity in both the density and pressure slices.
      Due to the larger opening angle (and hence, lower injected density), simulation Q36-v01-$\uptheta$30 produces both a wider jet beam and a wider flaring point.
      This impacts the density of the flow downstream from the flaring point, which is denser for simulation Q36-v01-$\uptheta$25.
      This feature is also present in the large-scale density maps at later times, particularly for the northern lobe.
      Along with hydrodynamic quantities, the $1.4\,\textrm{GHz}$ surface brightness is shown for the inner region of the radio source.
      The wide, conical jet structure is faintly visible in the surface brightness map, however, it is dominated by emission from particles contained within the cocoon.
      Narrower bright spots of emission are seen for the $25\degr$ half-opening angle jet.
      Finally, we note that the morphologies produced by these jets are lobed FR Is.
      At early times a bow shock is driven into the ambient medium by the jets.
      This bow shock works to contain the low-density jet material and leads to the formation of a cocoon through forward-flow, rather than backflow as in FR IIs.
      This also leads to compression of the lobe material near the tip of the lobe, producing a hotspot-like feature.
      The strength of the feature declines with time.
      While at early times, this can sometimes be the brightest feature, at late times, the source is consistently in the FR I regime.
      This is similar to what has been seen by \citet{Krause2012} with more limited methods.

      The distance at which the flaring point occurs is not static over the lifetime of the radio source.
      As these are lobed FR Is, the recollimation process is driven by the cocoon pressure, not the ambient medium \citep{Krause2012}.
      Therefore as the cocoon evolves the distance to the flaring point is also expected to evolve.
      This is shown in \cref{fig:cd_frI_flaring_region}, which plots the distance to the flaring point as a function of time for both the northern (solid line) and southern (dotted line) lobes.
      Half-opening angle does not have a significant effect on the evolution of flaring region distance over time, which peaks in the first $\sim 5\myr$ of the source lifetime, before decreasing as the source continues to evolve towards pressure equilibrium with the environment.

    \subsubsection{Observable signatures}
    \label{sec:cd_radio_observables}

      In \cref{fig:cd_frI_surface_brightness} surface brightness maps for both low power, slow jet simulations at $t=100\myr$ are shown.
      These surface brightness maps are calculated as for \cref{fig:cd_jet_population_surface_brightness}; the emissivity calculation and observing parameters are described in \cref{fig:cd_frI_surface_brightness}.
      Three different orientations are shown: in the plane of the sky, or inclined at $30\degr$ or $60 \degr$ with respect to the observer.
      While Doppler boosting is included in the emissivity calculations, the contribution is negligible given the low jet velocities in these simulations.
      This implies that for a given absolute inclination angle, whether the northern lobe is inclined towards or away from the observer should have little effect on the observed surface brightness.
      This is confirmed by the surface brightness maps: little to no surface brightness difference exists between the two lobes for a given observing angle.
      At this time, the source has long reached pressure equilibrium with its environment, so that the hotspot-like features near the tip of the lobes have vanished and a pure FR I structure has emerged.
      The effect of the environment is evident in both simulations, causing the southern lobe to curve with respect to the jet core and northern lobe; a direct reflection of the underlying gas pressure field.

      For all orientations at $100\myr$, distinct FR I features are present: a bright flaring region near the core, followed by plume-like emission.
      These FR Is are lobed; however, an edge-darkened structure is observed.
      If the surface brightness sensitivity was decreased, the observable size of the plume-like emission downstream of the flaring region would also decrease, in agreement with analytic models for FR I radio sources \citep{Turner2018a}.

      The evolution of $1.4\,\textrm{GHz}$ total luminosity with source size is shown in \cref{fig:cd_frI_pd_tracks}.
      There is no significant difference in luminosity evolution between the two simulations; however, the morphology is different, leading to slightly different length evolution.
      The source evolution in luminosity covers $\sim1.5$ dex over $100\myr$, before rapidly declining once the jet is switched off.

      The evolution of the FR index with time is presented in \cref{fig:cd_frI_fr_index}, for both edge-on and inclined $60\degr$ orientations.
      There are three distinct stages in the FR index evolution.
      The first is the initial jet injection phase, which lasts from $0$ to $\sim 20\myr$.
      In this phase, the FR index is dominated by whether the brightest point in a lobe is the hotspot or the flaring region.
      For sources in the plane of the sky, this changes often and so this region is noisy (c.f. \cref{fig:cd_frI_zoomed}).
      In contrast, the inclined sources consistently have FR indices in the FR I range as the hotspot appears recessed within the lobe, and there is no longer a separation between the flaring region and hotspot.

      The second phase lasts from $20$ to $100\myr$, during which time the FR index is purely within the FR I range.
      Within this phase, emission from the flaring region dominates, regardless of observing orientation, and the source is consistently identified as an FR I.
      The third and final phase starts when the jet is switched off at $t=100\myr$, and consists of the remnant evolution discussed below in \cref{sec:cd_frI_remnant}.

      The spectral index for both simulations are plotted as a function of frequency in \cref{fig:cd_frI_total_spectral_index}, and the corresponding integrated lobe spectra in \cref{fig:cd_frI_spectra}.
      At $t=100\myr$ when the jet is active, the spectral index shows very little evolution with frequency.
      While spectral steepening is expected to occur in expanding lobes continuously fed with recently accelerated electrons, it is not observed in the frequency range considered here.
      The spatial distribution at $t=100\myr$ of $\alpha^{1400}_{150}$ (low), $\alpha^{9000}_{1400}$ (high), and $\alpha^{9000}_{1400} - \alpha^{1400}_{150}$ (spectral curvature) are shown in \cref{fig:cd_frI_spectral_index_map_low,fig:cd_frI_spectral_index_map_high,fig:cd_frI_spectral_curvature_map} respectively, for the same three radio source orientations as in \cref{fig:cd_frI_surface_brightness}.
      The spectral index is constant within the flaring region, as expected for a population of recently accelerated electrons.
      It then steepens along the jet for both the low and high spectral indices (indicating older electrons), before flattening slightly at the lobe tips for the low spectral index, likely due to the concentration of recently shocked forward flowing electrons at these locations.
      A steep spectral index is observed within the lobes, due to the older population of electrons.
      This steepening is more pronounced in the high spectral index map, as predicted by the frequency dependence of radiative losses.

    \subsubsection{Remnant evolution}
    \label{sec:cd_frI_remnant}

      The low power simulations are switched off at $100\myr$ for both simulations, after which they are evolved for another $100\myr$ to study the remnant phase.
      After the jets switch off, the inflated low-density cocoons rise buoyantly away from the injection region and slowly morph into pancake-shaped bubbles, while a sound wave continues to propagate through the environment.
      The buoyant rise velocity is several hundred km s$^{-1}$, consistent with the results of \citet{Churazov2001}.
      This evolutionary stage is shown in \cref{fig:cd_frI_surface_brightness_remnant}, which plots the radio surface brightness for these two simulations, with the same limits and observing parameters as in \cref{fig:cd_frI_surface_brightness}.
      As the inflated jet cocoons rise, they depart from the jet axis of symmetry due to the environment dynamics and asymmetry.
      This bending is most evident in the northern lobe of Q36-v01-$\uptheta25$, and the southern lobe of Q36-v01-$\uptheta30$.
      Both simulations have different radio lobe morphology for the northern and southern lobes.

      Once the jet is switched off, the flaring region begins to rapidly fade as the population of newly-shocked electrons is not replenished.
      Meanwhile, the outer parts of the lobes also fade but at a slower rate, due to the delay between the jet switching off and the disappearance of forward flow downstream of the flaring region.
      We find significant spectral evolution in the remnant phase, as expected.
      The $t=130$ and $200\myr$ spectral index and integrated spectra in \cref{fig:cd_frI_total_spectral_index,fig:cd_frI_spectra} respectively demonstrate the frequency-dependent loss process; while electrons in the radio lobes are still emitting $30\myr$ after the jet switched off, the emission spectrum steepens significantly.
      This occurs similarly for both simulations.
      At $t=200\myr$, no electrons above some cut-off frequency between $1.4\,\textrm{GHz}$ and $5.5\,\textrm{GHz}$ are emitting.

\section{Discussion}
\label{sec:cd_discussion}

CosmoDRAGoN simulations explicitly connect small (kpc) scales on which jet collimation occurs to larger (tens to hundreds of kpc) scales relevant to maintenance-mode AGN feedback. By calculating synthetic radio emission in post-processing, we are able to explore the connection between jet injection, feedback, and emergent radio source characteristics. We discuss some key early results below.

\subsection*{Radio morphology}
\label{sec:cd_radio_morphology}

Whether radio source morphology is primarily determined by jet or environment properties is still an open question.
Some analytical \citep{Bicknell95} and numerical \citep{Perucho2014} theoretical work suggests that slow jets are more likely to form core-brightened FR I sources; mass-loading by either direct entrainment or stellar winds \citep[e.g.][]{WykesEA15,Laing2014} may sufficiently slow down the initially relativistic jets on sub-kpc scales.
\citet{Alexander06} and \citet{Krause2012}, on the other hand, have argued that the key parameter in determining radio source morphology is the jet opening angle: jets with sufficiently large angles will run out of forward ram pressure before the jet is collimated, producing FR I sources; while narrower jets with the same speed and kinetic power will produce FR II sources.
CosmoDRAGoN simulations contribute to this discussion by enabling a comparison of jets with different speeds, kinetic powers, and opening angles.
As \cref{fig:cd_frII_fr_index,fig:cd_frI_fr_index} show, fast, narrow jets (the Q38-v98-$\uptheta$7.5 simulation in \cref{fig:cd_frII_fr_index}) retain their FR II morphology to hundreds of kpc sizes.
Wider jets transition to FR I morphology (compare Q38-v98-$\uptheta$7.5 and Q38-v98-$\uptheta$25 simulations in \cref{fig:cd_frII_fr_index}), and this transition happens earlier at lower jet powers (compare Q36-v30-$\uptheta$25 and Q38-v30-$\uptheta$25 simulations in \cref{fig:cd_frII_fr_index}) as expected, because the low-power jets reach pressure equilibrium with their environment first.
We find that even fast jets (Q38-v98-$\uptheta$25 simulation) make the transition to FR I morphology at sufficiently late times; however, this transition occurs earlier for slow jets, and on smaller spatial scales for jets with low kinetic power (\cref{fig:cd_frI_fr_index}).
We therefore expect even the most powerful jets to eventually form FR I lobes if they have a sufficiently wide opening angle; ongoing high surface brightness sensitivity surveys such as LOFAR LoTSS \citep{ShimwellEA19,ShimwellEA22}, ASKAP EMU \citep{NorrisEA21}, the VLA Sky Survey \citep{LacyEA20} and Meerkat MIGHTEE \citep{JarvisEA16} will detect sufficiently large numbers of Giant Radio Galaxies to test this prediction.

\cref{fig:cd_jet_population_density} shows that the details of jet energy injection are important for determining the observed bow shock and radio source morphology.
Narrow jets (Q38-v98-$\uptheta$7.5) are collimated earlier and produce more elongated cocoons; but at fixed opening angle fast, light, jets are better at isotropising feedback (cf Q38-v98-$\uptheta$25 and Q38-v30-$\uptheta$25 simulations).
This result was previously reported by \citet{Krause05} and \citet{PeruchoEA17}.
Hence, relativistic jets must be modelled properly to accurately represent jet feedback -- a major challenge for current cosmological galaxy formation models. 
We defer to future work detailed investigations of the relationship between jet parameters, radio source morphology, and feedback efficiency.

\subsection*{The role of environment}
\label{sec:environment}

In this paper we have presented early CosmoDRAGoN results focused on a single environment, representative of a low-redshift cluster.
Our full simulation suite will cover a broad range of environments, including galaxy groups and clusters, at several cosmic epochs.
It is well established that environment plays an important role in radio source dynamics and propagation: at a given age, the same jet pair will produce a more compact, more luminous radio source in a denser environment \citep{BegelmanCioffi89,KaiserEA97,Shabala2008,HardcastleKrause2013,ShabalaGodfrey13,TurnerShabala15,Hardcastle18}.
The importance of large-scale environmental dynamics (i.e. ``cluster weather'') depends on both jet and environment properties: \citet{Krause2012} define a length scale $L_2 = \left( \frac{Q_{\rm j}}{\rho_{\rm x} c_{\rm x}^{3}} \right)^{1/2}$ at which the radio source comes into approximate pressure equilibrium with its surroundings; here $Q_{\rm j}$ is the jet kinetic power, and $\rho_{\rm x}$ and $c_{\rm x}$ are the environment density and sound speed, respectively.
The dynamics of sources smaller than $L_2$ is dominated by the jet momentum flux; sources much larger than $L_2$ are in the buoyant regime, and hence susceptible to large-scale gas motions.
Because $L_2$ is smaller in denser environments and for low power jets, this effect will be most pronounced for low power jets in clusters.
Once the jets switch off, the source enters a remnant phase; this phase is characterised by a markedly slower expansion \citep[e.g.][]{KaiserEA03,YatesEA18}.
\cref{fig:cd_frI_surface_brightness,fig:cd_frI_surface_brightness_remnant} show clearly that remnant FR I lobes are affected by cluster dynamics, becoming increasingly asymmetric with time.
Such asymmetry may pose a challenge for accurately identifying lobe pairs in remnant radio sources \citep{Brienza2017,MahatmaEA18,JurlinEA20} and subsequent interpretation of observed remnant populations \citep{GodfreyEA17,ShabalaEA20}. We defer detailed exploration of these questions to a future paper. 

\section{Summary and conclusions}
\label{sec:cd_summary}

In this paper, we have presented an overview of the CosmoDRAGoN project: the first suite of simulations of conical, relativistic and non-relativistic jets in cosmological environments derived from galaxy formation simulations.
By exploring a wide range of jet parameters (kinetic power, opening angle, speed) and environments, we are able to study the effects of environment on observable jet and radio lobe properties, as well as the feedback imparted on the circum- and intergalactic gas.
Our simulations have sufficient resolution to resolve collimation of the initially conical jets, and their subsequent propagation to scales of hundreds of kiloparsecs -- scales characteristic of evolved radio galaxies, and important for maintenance-mode feedback.
Jets are evolved in both active and remnant phases, then post-processed using a semi-analytic framework to calculate synthetic synchrotron emission, including spatially resolved radio spectra.
This approach enables a direct comparison with radio observations, for the first time connecting radio observables and feedback in cosmological environments across a broad parameter space.

We have described the technical details underpinning these simulations, including the selection and interpolation of initial conditions; the jet injection method; and environment stability.
We have presented an overview of the data products, and our post-processing pipeline that yields synthetic radio observables in addition to fluid quantities relevant for studying jet feedback.

Drawing on six representative simulations, we have explored the evolution of high and low power jets, injected into cosmological environments with a range of speeds and opening angles.
Our simulations produce observational features typical of real radio sources, including both core- (FR I) and edge-brightened (FR II) morphologies, and complex surface brightness distributions and radio spectra.
We confirm earlier findings from simulations of jets in idealised environments that jet injection parameters play a key role in determining the resultant radio source morphology; this result has important implications for implementations of jet feedback in cosmological simulations.
We also find that cluster weather can significantly affect radio source morphology, particularly for low power jets at late times.
By spanning a wide range of jet and environment properties, the full CosmoDRAGoN simulation suite will provide insights into the complex relationship between AGN jets and their environments, and provide a framework for connecting the observed radio source populations to physical mechanisms responsible for jet triggering and feedback.
We will report on these findings in forthcoming papers.

  \begin{acknowledgements} 

  We thank an anonymous referee for their useful comments.
  This work was supported by the Australasian Leadership Computing Grants scheme, with computational resources provided by NCI Australia, an NCRIS enabled capability supported by the Australian Government.
  We also gratefully thank the Tasmanian Partnership for Advanced Computing (TPAC) of the University of Tasmania for the computational resources provided.
  This work has made use of data from \textsc{the three hundred} collaboration (\url{https://www.the300-project.org}) which benefits from financial support of the European Union’s Horizon 2020 Research and Innovation programme under the Marie Skłodowskaw-Curie grant agreement number 734374, i.e. the LACEGAL project.
  \textsc{the three hundred} simulations used in this paper have been performed in the MareNostrum Supercomputer at the Barcelona Supercomputing Center, thanks to CPU time granted by the Red Espa\~{n}ola de Supercomputaci\'{o}n.
  PYJ thanks the University of Tasmania for an Australian Postgraduate Award and the ARC Centre of Excellence for All Sky Astrophysics in 3 Dimensions (ASTRO 3D; CE170100013) for a stipend. CP acknowledges the support of ASTRO 3D. 
  We acknowledge the work and support of the developers providing the following Python packages: Astropy \citep{AstropyCollaboration2018,AstropyCollaboration2013}, JupyterLab \citep{Jupyter}, Matplotlib \citep{Matplotlib}, NumPy \citep{NumPy}, SciPy \citep{SciPy}, and yt \citep{Turk2011}.
  We thank Attila Juhasz for their work on sphtool.

  \end{acknowledgements}

\section*{Data Availability}
  The CosmoDRAGON raw simulation data were generated at NCI Australia and TPAC.
  Derived data supporting the findings of this study are available from the corresponding author, PYJ, upon reasonable request.

  \begin{appendix}

  \end{appendix}

  \bibliographystyle{pasa-mnras}
  \bibliography{cosmo-dragon.bib}

\end{document}